\documentclass[12pt]{iopart}

\usepackage{url}
\usepackage{amsmath}
\usepackage{graphicx}
\usepackage{multirow}
\usepackage[backend=biber, 
            style=numeric,          
            maxnames=10, minnames=10,maxbibnames=10, minbibnames=10,            
            doi=false,              
            url=false,              
            isbn=false,             
            giveninits=true,        
            natbib=true,            
            sortcites=true,         
            block=space,
            sorting=none]{biblatex}
\DeclareFieldFormat[article]{title}{#1}                
\DeclareFieldFormat[article]{journaltitle}{\mkbibemph{#1}} 

\DeclareFieldFormat{pages}{#1} 

\DeclareBibliographyDriver{article}{%
  \printnames{author}                 
  \addspace
  \printfield{year}                   
  \addspace
  \printfield[journaltitle]{journaltitle} 
  \addspace
  \iffieldundef{volume}{}{\textbf{\printfield{volume}}} 
  \addspace
  \printfield{pages}                  
}

\usepackage[normalem]{ulem}
\usepackage{xcolor}

\begin{document}

\title[Improving early detection of gravitational waves from binary neutron stars]{Improving early detection of gravitational waves from binary neutron stars using CNNs and FPGAs}

\author{Ana Martins$^{1}$, Melissa Lopez$^{2, 3}$, Gregory Baltus$^{4}$, Quirijn Meijer$^{2, 3}$, Marc van der Sluys\footnote{The research leading to these results has received funding from the European Union's Horizon 2020 Programme under the AHEAD2020 project (grant agreement No. 871158). This publication is part of the project Cortex with project number 00686766 of the research programme NWA which is (partly) financed by the Dutch Research Council (NWO).}$^{2, 3}$, Chris Van Den Broeck$^{2, 3}$, Sarah Caudill\footnote{S.C is supported by the National Science Foundation under Grant No. PHY-2309332.}$^{5, 6}$}
\address{$^{1}$ Institute of Theoretical Astrophysics, University of Oslo, Oslo, Norway}
\address{$^{2}$ Institute for Gravitational and Subatomic Physics (GRASP), Utrecht University}
\address{$^{3}$ 
Nikhef -- National Institute for Subatomic Physics, Netherlands\\}
\address{$^{4}$ STAR Institut, Université de Liège, Liège, Belgium}
\address{$^{5}$ Department of Physics, University of Massachusetts Dartmouth, Center for Scientific Computing}
\address{$^{6}$ Data Science Research, University of Massachusetts Dartmouth, Dartmouth, USA}

\ead{a.i.s.martins@astro.uio.no}
\vspace{10pt}
\begin{indented}
\item[]\today
\end{indented}

\begin{abstract}
The detection of gravitational waves (GWs) from binary neutron stars (BNSs) with possible telescope follow-ups opens a window to ground-breaking discoveries in the field of multi-messenger astronomy. 
With the improved sensitivity of current and future GW detectors, more BNS detections are expected in the future. Therefore, enhancing low-latency GW search algorithms to achieve rapid speed, high accuracy, and low computational cost is essential. One innovative solution to reduce latency is the use of machine learning (ML) methods embedded in field-programmable gate arrays (FPGAs).

{In this work, we present a novel \texttt{WaveNet}-based method, leveraging the state-of-the-art ML model, to produce early-warning alerts for BNS systems. Using simulated GW signals embedded in Gaussian noise from the Advanced LIGO and Advanced Virgo detectors' third observing run (O3) as a proof-of-concept dataset, we demonstrate significant performance improvements. Compared to the current leading ML-based early-warning system, our approach enhances detection accuracy from 66.81\% to 76.22\% at a 1\% false alarm probability. Furthermore, we evaluate the time, energy, and economical cost of our model across CPU, GPU, and FPGA platforms, showcasing its potential for deployment in real-time gravitational wave detection pipelines.}
\end{abstract}

%
%
%
%
%

\section{Introduction}

In recent years, Astrophysics and adjacent sciences have been expanding at an astronomical pace. A significant driver of this growth is the detection of gravitational waves (GW{s}), which are ripples in spacetime generated by astronomical cataclysms. 
In September 2015, the Laser Interferometer Gravitational-Wave Observatory (LIGO) \cite{LIGOScientific:2014pky} and Virgo \cite{ VIRGO:2014yos} collaborations confirmed the existence of GW{s} {by} detecting a binary black hole (BBH) merger \cite{PhysRevLett.116.061102}. Since then, over 90 confident events have been detected in the past three observation runs by LIGO-Virgo collaboration \cite{LIGOScientific:2018mvr, LIGOScientific:2020ibl, LIGOScientific:2021usb, KAGRA:2021vkt}. {Among these astronomical events, only two binary neutron star (BNS) mergers were detected \cite{LIGOScientific:2017vwq}. }



{As GW detectors become increasingly sensitive with each upgrade, the enhanced capabilities of second-generation detectors and the future third-generation detectors,} such as the Einstein Telescope (ET) \cite{Punturo:2010zz}{,} Cosmic Explorer \cite{Reitze:2019iox} {or the space-based Large Interferometer Space Antenna \cite{amaroseoane2017laserinterferometerspaceantenna}}{, will allow for more frequent detection of binary neutron star (BNS) mergers and other exotic systems.} Nonetheless, this increase in sensitivity comes at {increase in computational complexity}, as more sources will be detected. In the era of ET{,} it is estimated that $8 \times 10^4\, $yr$^{-1}$ BBH  \cite{Iacovelli:2022bbs} and $7 \times 10^4\, $yr$^{-1}$ BNS \cite{Kalogera:2021bya} will be detected, which could lead to over 400 daily GW merger events, as well as the detection of other exotic sources \cite{Baiotti:2005vi}. {Moreover, while {BNS} GW signals are detectable for seconds with current detectors, they will be present for hours in the ET era.}

The current state-of-the-art for detecting modelled GW signals, known as matched-filtering—a technique based on cross-correlating models, or templates, with GW detector data \cite{Messick:2016aqy, Sachdev:2019vvd, Tsukada:2023edh, usman2016pycbc, Allen:2005fk}—has led the LIGO-Virgo-KAGRA collaboration to develop several low-latency matched-filtering pipelines for producing real-time GW alerts \cite{Sachdev:2020lfd, Nitz_2020, Magee:2021xdx, Kovalam_2022, Chaudhary:2023vec}. While these methods have been successful, matched-filtering techniques are known to be computationally intensive, posing challenges for future detectors and {negatively affecting the environment \cite{andre2018bigdata}}.

In recent years, machine learning (ML) techniques have gained significant interest due to their success in various tasks and domains {\cite{Aveiro_2022}}. A key advantage of ML is its fast inference, with most computations occurring during the training stage. This is crucial for GW searches, particularly for early warning of BNS GW signals in multi-messenger astrophysics. The goal is to detect GW signals during the inspiraling phase, where both neutron stars are orbiting around each other, {before they merge in{to} a {more massive} astronomical object (known as merger phase)}, to enable {electromagnetic telescope follow-ups}. {This task is challenging because GW signals are often buried in detector noise, with the signal being weaker during the inspiral phase and strengthening as it approaches the merger.} 

GPUs, widely used for deploying ML algorithms, are well-known for being costly, consuming large amounts of energy, having short lifetimes and potentially introducing high latencies within data transfer. In a low-latency framework, these characteristics could hamper the early detection of GW signals. An interesting alternative is to embed ML algorithms in field-programmable gate arrays (FPGAs){. FPGAs show promise by having longer lifetimes and being more energy-efficient than state-of-the-art hardware. Furthermore, high-end FPGAs are faster than CPUs and GPUs \cite{6589302}, and low-end FPGAs are more affordable than CPUs or GPUs in the same tier. We choose the latter to explore their applicability and limitations.}

{In this work}, to build more sustainable and environmentally-friendly alternatives, we explore the performance of FPGAs with respect to GPUs and CPUs to detect the early inspiral of BNS GW signals with ML-based algorithms\footnote{The code is accessible in GitHub: \url{https://github.com/anaismartins/deploying-custom-nn-in-fpga}}.
This paper is structured as follows{:} in Section \ref{sec:SOTA}, we address the state of the art in early warning with ML-based methods and motivate the need to improve the precision at a limited computational cost; in Section \ref{sec:dataset}, we describe the input data; in section \ref{sec:methodology}{,} {we describe the hardware employed and ML architectures}, as well as learning and quantization strategies {(reduction from floating-point to fixed-point representation)}; in Section \ref{sec:results}{,} we present the main findings of this work, comparing our model with the state of the art in terms of accuracy and computational cost; finally, Section \ref{sec:conclusions} presents the conclusions of this research with avenues for future research.

\section{{Related work}}\label{sec:SOTA}

The challenges in GW research require innovative solutions, and ML has emerged as a crucial tool for addressing them due to its adaptability and transversality. 
In the past few years, researchers have explored different ML applications to GW data analysis, such as detection of modelled \cite{Baltus:2021nme, Baltus:2022pep} and unmodelled GW signals \cite{2020arXiv200204591C, LopezPortilla:2020odz, Lopez:2021rci, Meijer:2023yhn, Boudart:2022xib, Boudart:2022apz}, non-transient burst noise characterization \cite{bahaadini2018machine, Laguarta:2023evo} and synthetic data generation \cite{Lopez:2022lkd, Lopez:2022dho, Dooney:2022arh, Dooney:2024pvt}, among others. Refer to \cite{Cuoco:2020ogp} for a review.

Traditional GW searches use the inspiral part of the waveform, as early efforts have shown that it is possible to detect a signal with only a fraction of the waveform model with matched-filtering techniques \cite{Sachdev:2020lfd}. The first early warning ML-based algorithm implemented \texttt{ResNet50} \cite{2015arXiv151203385H} training on spectrograms containing solely the inspiral part of the waveform \cite{Wei:2020sfz}. However, as noted in \cite{Baltus:2021emh}, this added pre-processing step slows the inference by $\sim 0.5\,$s.
To address this, \cite{Baltus:2021emh} suggested using GW detector time series in design sensitivity, and employing 1-dimensional convolutional neural networks (CNN{s}), eliminating the need for this pre-processing step and enhancing efficiency. This proof of concept work led to subsequent efforts in this line of research \cite{Wei:2020xrl,  Baltus:2021nme, Yu:2021vvm, Baltus:2022pep, Alfaidi:2024ioo, vanStraalen:2024xiq}. 

In particular, \cite{Baltus:2022pep}, building on the previous work of \cite{Baltus:2021emh} and \cite{Baltus:2021nme}, extended this method to the simulated noise of third and fourth observing runs (O3 and O4), as well as real O3 data. While \cite{Baltus:2022pep} is the current state of the art, some of the limitations of this study are described below:

\begin{itemize}
\item \textbf{Large number of trainable parameters:} The fixed input-signal of this work has a duration of $300\,$s with a sampling frequency of $512\,$Hz, resulting in $155,648\, $data points. Due to the long size of the input, large and complex models, that could lead to higher accuracy, would not fit into the device's memory. 
\item \textbf{{Catastrophic forgetting}:} This investigation successfully implements curriculum learning --where the model learns easier examples to then transition to harder ones-- by lowering the maximum frequency of the signals or rather detecting the signal earlier. However, it is observed that the model forgets easier steps of curriculum learning due to its over-fitting. 
\item \textbf{{High false positive rate:}} The model has a {large} number of false positives, i.e. misclassifying detector noise as GW signal, regardless of its decision threshold.
\end{itemize}

To overcome the limitations of the previous work, hereafter referred to as \texttt{FindCNN}, we implemented an {algorithm inspired by} \texttt{WaveNet} \cite{2016arXiv160903499V} named \texttt{GWaveNet}. Similarly to \cite{Wei_2021} and \cite{Meijer:2023yhn}, this choice was motivated by the similitude between GW time series data and audio or speech signals. For a fair comparison, we reproduced \texttt{FindCNN} and trained and tested both models on the same simulated GW detector data. 

FPGAs have been used for ML applications \cite{fpga-nns-paper}, also in high-energy physics  \cite{khoda2022ultralow, Que_2024}. We highlight the pioneering work in \cite{Que:2021cqo} for the implementation of recurrent neural networks for anomaly detection in GW. 
As added value, in this work we evaluate the performance  of {the \texttt{FindCNN}} and \texttt{GWaveNet} models in CPU, GPU and FPGA {devices}, as well as their time, energy consumption and economical cost.

\section{Data set and pre-processing}
\label{sec:dataset}

A GW search algorithm processes raw detector data, producing {{a score about} whether} it contains a  GW signal. {We {can find GWs using}  }supervised {learning}, where we distinguish detector noise (called a negative class) from detector noise plus GW signal (called a positive class).

In a real scenario, the negative class is much larger than the positive class, as {only} {$\sim100$ GW signals} have been discovered at the present date. Thus, it is required to simulate GW signals, adding them to detector noise. This process is commonly referred to as ``injections", and they will be the positive class of this study. Hence, we construct a balanced data set of samples {with only detector noise (negative class), and injections (positive class)}. Simulated Gaussian noise is generated with the average power spectral density of the third observing run (O3) for Hanford, Livingston and Virgo\footnote{We use the PyCBC package \cite{Biwer_2019}. In particular, the power spectral densities \texttt{aLIGOaLIGO140MpcT1800545} and \texttt{aLIGOAdVO3LowT1800545}, an average of O3, {respectively for the Advanced LIGO and Advanced Virgo detectors \cite{KAGRA:2023pio}. Refer to \url{https://gwosc.org/O3/o3_details/} for technical details of the observing run.}}.

As proposed in~\cite{Baltus:2022pep}, to encompass all potential BNS systems, the component masses are uniformly distributed $\in [1, 3]\,\text{M}_{\odot}$, and the waveform approximant \textit{SpinTaylorT4} \cite{Buonanno:2002fy} {(used to model inspirals)} was chosen. Moreover, the sources are uniformly distributed over the sky, including spin effects.

\begin{figure}[h]
\centering
\includegraphics[width=0.9\textwidth]{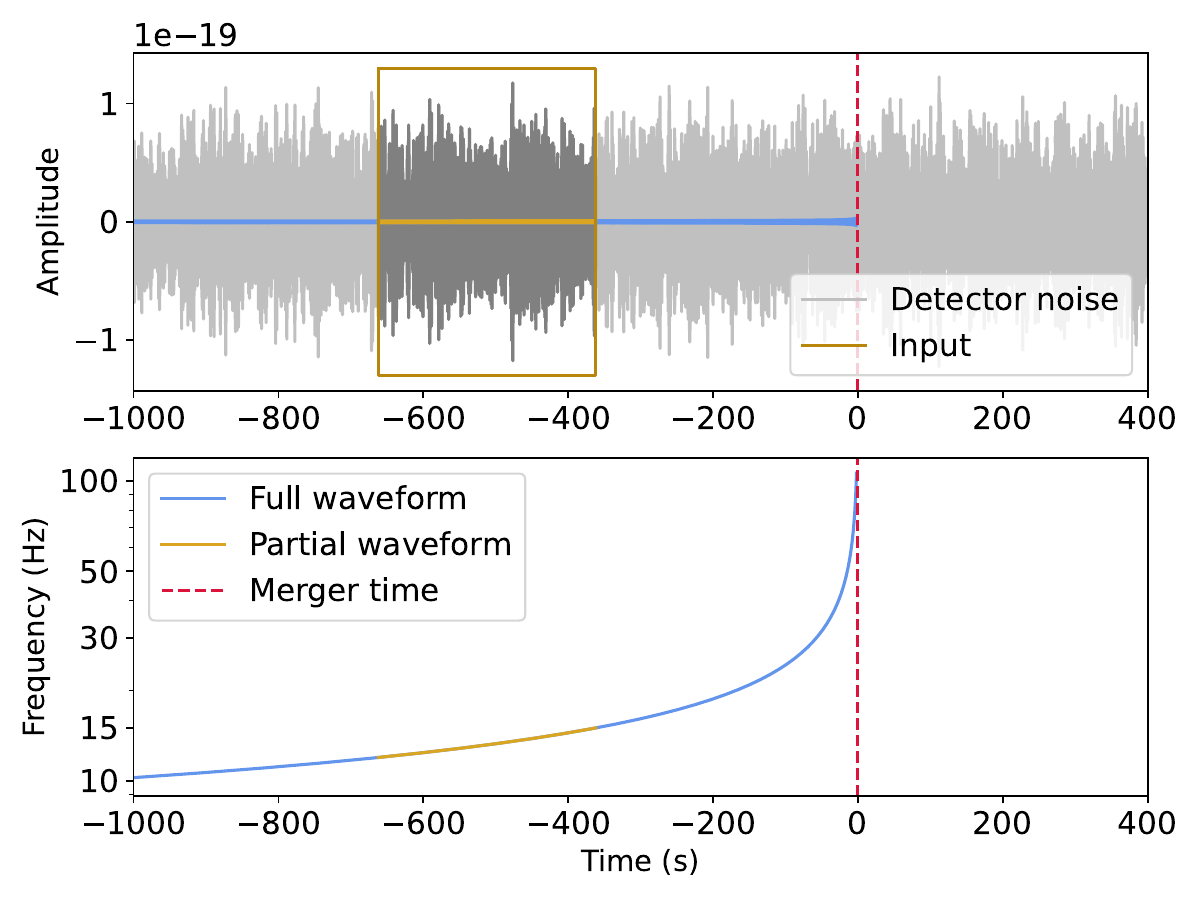}
\caption{\textit{(Top)} Representation of {the inspiral phase, where both neutron stars are orbiting around each other,  of a} GW signal (blue) buried in simulated raw LIGO noise (grey){. The vertical red line indicates the merger time}. The GW waveform was modelled using \textit{SpinTaylorT4}-{which only models the inspiral phase}-with progenitor masses ($m_{1} =1.46\,\text{M}_{\odot}, m_{2} = 1.27\,\text{M}_{\odot}$), similar to GW170817. The input to the ML model is contained within the yellow rectangle{, i.e. $300\,$s of data sampled at $512\,$Hz}. \textit{(Bottom)} Frequency evolution of the full GW waveform (blue) and the partial waveform (yellow).}
\label{fig:input}
\end{figure}

As GWs have {a wea{k}} amplitude, they are {hard to distinguish from} detector noise. For illustration, in Fig. \ref{fig:input} (top panel) we show the GW inspiral of a signal similar to GW170817 \cite{LIGOScientific:2017vwq} in raw Advanced LIGO detector noise. To highlight its amplitude a common practice is to {filter the data between $10$ and $100\,$Hz, and} whiten it, i.e. make the signal  Gaussian-like with uniform variance by removing all the correlation of the noise \cite{Akhshi2020ATA}. Afterwards, we normalize the data {$\in [-1, 1]$}.
As we can observe in Fig. \ref{fig:input}, the inspiral phase lasts several minutes, so{,} as in \cite{Baltus:2022pep}{,} we pre-select $300\,$s of data. In{ the} bottom panel{,} we show the relation between time and frequency of the source. Because of this relation, the difficulty of the input is governed by the maximum frequency $f_{max}$ {within a time window}. Samples with higher (lower) $f_{max}$ will be closer (further) to the merger, meaning that they will be detected later (earlier). 

As the Hanford, Livingston and Virgo observatories measure GW signals independently, {a signal that appears simultaneously in data from multiple detectors {at the same time}} is more likely to be of astronomical origin. From an ML perspective, independent detector data is input as separate channels.

\section{Methodology}
\label{sec:methodology}

{In this work, we} do not only aim to detect GW signals as early as possible, going down in frequency $f_{max}$, but also to minimize the number of false positives (FP), to avoid sending false alerts to electromagnetic telescopes.
Thus, we have developed a WaveNet-like architecture, \texttt{GWaveNet}, that we compare to the current state of the art, \texttt{FindCNN}. In the following sections{,} {we describe the hardware employed for the deployment of the models and provide an overview of \texttt{FindCNN} and  \texttt{GWaveNet}}. 

\subsection{Hardware specifications}
\label{sec:hardware}
The manufacturing and use of state-of-the-art hardware are causing substantial damage to the environment and this is only predicted to increase, with {most emissions from data-center equipment being related to manufacturing \cite{9407142}} and computation energy predicted to hit the world's energy production capacity by 2040 \cite{mitnews_carbonfootprint}. In this context, it is of interest to understand the computational cost of low-latency algorithms of large physics experiments, such as Advanced LIGO-Virgo detectors, as well as their lifetime under heavy usage. In the following, we provide the specifications of the hardware employed in this work.

\begin{itemize}
\item[-] \textbf{CPU:} We use the AMD EPYC 7551P 32-Core Processor \cite{techpowerup_epyc7551p}, {which} has an expected lifetime of around 5 to 10 years. We use{d the full} 32 cores for testing.
\item[-] \textbf{GPU:} We used the NVIDIA Tesla V100 \cite{microway_tesla_v100} for training and validation, and NVIDIA GeForce GTX 1080 \cite{nvidia_gtx_1080} for testing and energy consumption calculation. Their expected lifetime is around 3 to 5 years. 
\item[-] \textbf{FPGA:} We use the AMD Kria KV260 Vision AI Starter Kit \cite{amdKriaKV260}. Its expected lifetime {is} around 10 to 15 years. We use the most recent {and fastest pre-made architecture for this system-on-chip}, B4096, with 4096 peak operations per cycle. Furthermore, we use 10 threads for testing{, which we found to be the optimal amount}.
\end{itemize}

\begin{figure*}[h!]
  \centering  
  \includegraphics[width=0.9\textwidth]{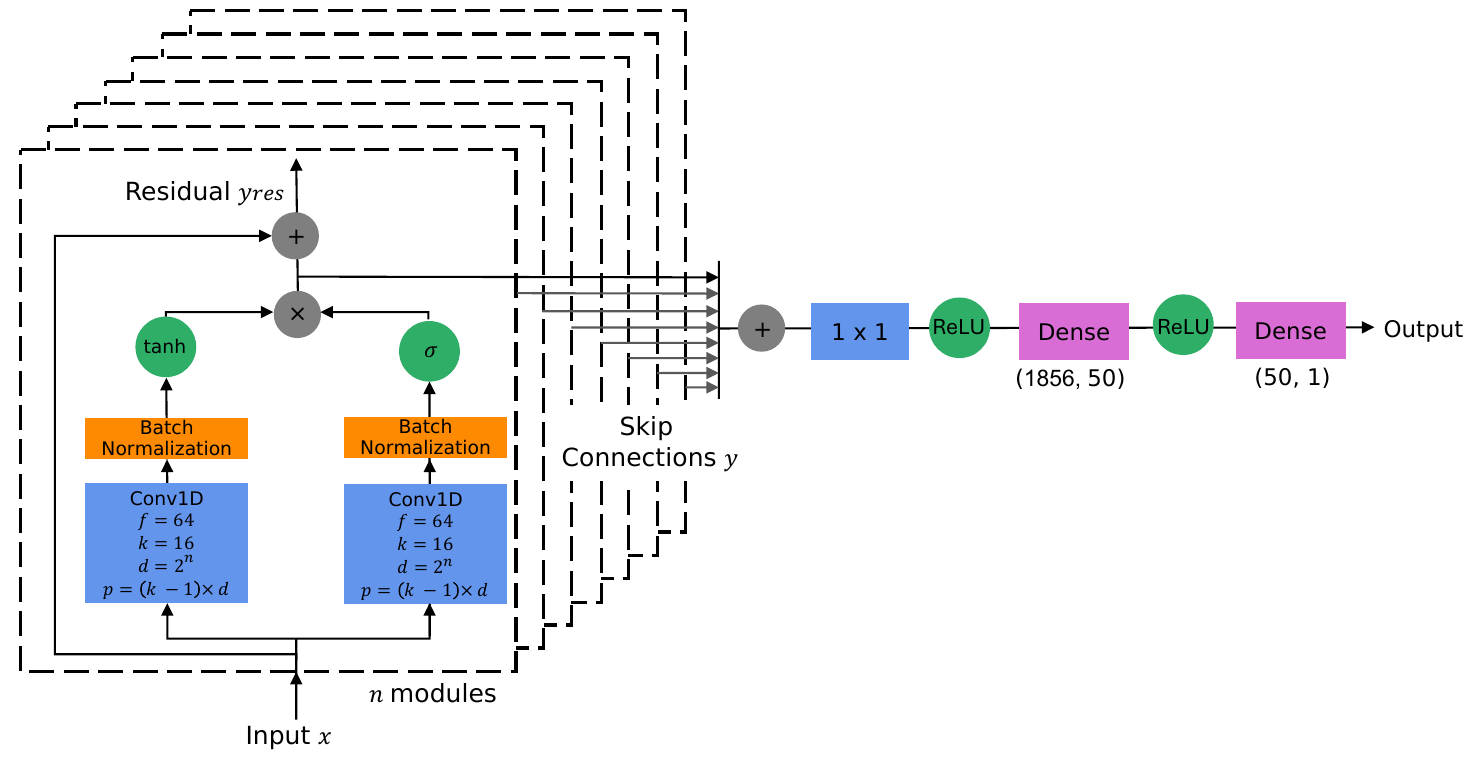}
  \caption{Overview of the modified \texttt{WaveNet} convolutional architecture for a single data stream, adapted from \cite{oord2016wavenet}. We present $n=7$ 1-dimensional \texttt{WaveNet} modules composed of dilated causal convolutions (blue) with the number of filters $f$, kernel size $k$, dilation $d$ and padding $p$, batch normalization layers (orange), $\tanh$ and sigmoid $\sigma$ activations (green). Furthermore, we show $1\times 1$ convolutions (blue), dense layers (pink), and ReLU and Softmax activations (green).}
  \label{fig:modulewavenet}
\end{figure*}

At the time of {their} launch{es}, the CPU and GPU {used for inference} were {nine} and {three} times more expensive than the FPGA used in this study. As of the current date of this investigation, the NVIDIA GeForce GTX 1080 GPU is {eight} years old, and the AMD EPYC 7551P is {seven} years old, with both considered obsolete. Typically, CPUs and GPUs become outdated even before their lifespan ends, as they struggle to compete with newer devices. In contrast, FPGAs are customizable, allowing modifications to keep them competitive throughout their lifetime. Consequently, the carbon footprint of using FPGAs is significantly lower than that of traditional processing units, not only because they consume less energy during operation but also due to their extended effective lifespan.

\subsection{Basic components}

\subsubsection{Common details}
We implemented \texttt{FindCNN} and \texttt{GWaveNet} using the \texttt{PyTorch} package \cite{paszke2019pytorch}. Both networks minimize the binary cross-entropy (BCE) loss in PyTorch, the \textit{BCEWithLogitsLoss} {i.e., Sigmoid activation integrated with BCE loss} \cite{pytorchBCEWithLogitsLoss}. We decided to employ this loss instead of \textit{BCELoss} since it is more numerically stable. Note that the original \texttt{FindCNN} \cite{Baltus:2022pep} uses \textit{BCELoss} instead. 

\subsubsection{\texttt{FindCNN}} It is a 1-dimensional CNN  \cite{Baltus:2022pep} that starts with a batch normalization layer \cite{ioffe2015batch}, to stabilize and accelerate the training process, followed by 5 convolutional blocks and two dense layers. The convolutional blocks are composed of a convolutional layer with stride 4, a ReLU activation function \cite{10.5555/3104322.3104425}, and a MaxPooling layer \cite{pytorchMaxPool1d} with stride 1. Each convolutional layer has successively kernel sizes $k$ $[16, 8, 4, 8, 16]$, and number of filters $f$ $[32, 64, 128, 256, 256]$. This structure allows capturing high-level features at earlier layers while capturing finer details at intermediate layers. Furthermore, the combination of large kernel $k$ and large filters allowed the last layers to capture high-level context with great expressiveness. Despite the compression of the input due to the usage of large kernel sizes $k$, the first dense layer is one of the most expensive ones, as it further compresses an input of $37,632$ data points to only $128$. This model has $6,179,303$ trainable parameters{, mainly due to the length of the input and large dense layers}.

\subsubsection{\texttt{GWaveNet}}
{\texttt{WaveNet} is an expressive CNN designed for the generation of high-fidelity speech audio \cite{oord2016wavenet}. The architecture is capable of handling long-range temporal dependencies, a characteristic desirable for GW data analysis, and in particular for BNS searches{, due to their long inspirals}. WaveNet-like models have been previously implemented in the field of GW searches \cite{Wei_2021, Meijer:2023yhn}. These implementations modify \texttt{WaveNet} for binary classification. This is a common point in the current investigation. However, previous \texttt{WaveNet} models were constructed as an ensemble with non-causal dilated convolutions, while we encode the information of all detectors in a single model with causal dilated convolutions.}

Causal dilated convolutions combine the power of dilation \cite{oord2016wavenet}, which allows {the NN} to capture a larger context  without increasing the computational load significantly, and causal convolutions, to effectively learn the time ordering {of the data}. {Our experiments show{ed} an improved performance when obeying the arrow of time with causal convolutions.} These convolutions are encapsulated in \texttt{WaveNet} modules.

We present the {full architecture of \texttt{GWaveNet}, including the} \texttt{WaveNet} modules in Fig. \ref{fig:modulewavenet}. We can observe how an input $x$ gets duplicated and each copy gets fed into a causal convolution block (blue), followed by a batch normalization layer (orange). Then, one output passes a $\tanh$ activation function (green), while the other passes a Sigmoid $\sigma$ activation function (green). This is known as gated activation \cite{2016arXiv160605328V}, {designed for} capturing complex relations in sequential data. Afterwards, the two outputs are multiplied (grey), generating the output of the module, $y$. At this point, the initial input $x$ is added to the output $y$ to form the residual input $y_{res}$ that passes to the next module. On the other hand, the output $y$ is sent towards the final output of the network that will feed the \textit{BCEWithLogitsLoss}. Before this step, all the $y$ outputs of the modules are summed, passing a $1\times1$ convolution (blue) and two dense layers (pink) with ReLU activation functions (green) \cite{10.5555/3104322.3104425}. It is relevant to note that original \texttt{WaveNet} modules used $1 \times 1$ convolutions {in between each module} for further compression, but due to the fixed number of filters $f$, this was not needed.

\texttt{GWaveNet} is composed of 7 \texttt{WaveNet} modules with kernels $k =16$, number of filters $f=64$ and dilation $d = 2^i$ for {$i \in [0, \dots, 6]$}. Because of this, the modules are highly expressive, having large receptive fields, which allows the compression of the input from $155,648$ to {$2,496$} data points and $64$ filters. Using a $1\times 1$ {convolutional} layer allows further compression of the first dense layer: now it needs to compress an input of {$2,496$} to $50$ data points. Thus, the number of trainable parameters of \texttt{GWaveNet} is $522,598$, $12$ times fewer parameters than \texttt{FindCNN}.

\subsection{Learning strategy}
\label{sec:learningstrategy}

As we mentioned in Sections \ref{sec:SOTA} and \ref{sec:dataset}, \cite{Baltus:2022pep} implemented a curriculum learning strategy going down in maximum frequency $f_{max}$. For this aim, five data sets with different average $f_{max}$ are built (see Table \ref{tab:frequencies}). In this way, the model first learns easier examples, closer to the merger, and slowly transitions to harder examples, further from the merger. Each data set is composed of $22,600$ samples for both classes, such that for each curriculum learning step we use $71\%$ for training, $9\%$ for validation and $20\%$ for testing. 

In GW data analysis, it is common to measure the loudness of the signal in terms of signal-to-noise ratio (SNR) \cite{Allen:2005fk}. However, \cite{Baltus:2021nme} and \cite{Baltus:2022pep} introduced the concept of partial {inspiral} signal-to-noise ratio (PISNR) to describe the corresponding SNR of the partial waveform{, seen by the CNN}. Thus, we use this magnitude for comparison (Table \ref{tab:frequencies}).

\begin{table}[h]
\centering
\begin{tabular}{|c|c|c|c|c|}
\hline
\bf{Data set} & \bf{$f_{min}\,$ (Hz)} & \bf{$f_{max}\,$ (Hz)} & \bf{SNR} & \bf{PISNR} \\ \hline
$D_1$        & 12.9                   & 40.0   & 100.3& 34.5               \\
$D_2$        & 12.8                   & 35.0   & 127.2 &  342               \\
$D_3$        & 12.6                   & 30.0   & 186.5&  34.7               \\
$D_4$        & 12.3                   & 25.0   & 344.6 & 34.7               \\
$D_5$        & 11.7                   & 20.0    & 1161.4& 34.6              \\ \hline
\end{tabular}
\caption{{In this classification task the positive class is a GW simulated signal added in raw detector noise. To enhance the learning we create five curriculum learning data sets $D$. In this table, we present the most relevant magnitudes of the GW simulations for each curriculum learning data set:} {the} average{s of the} minimum frequency $f_{min}$, {the} maximum frequency $f_{max}$, {the} network (Hanford, Livingston and Virgo) SNR and {the} PISNR.}\label{tab:frequencies}
\end{table}

To avoid {catastrophic forgetting}, we use a progressive curriculum learning approach, where the training set $\mathcal{D}_{T}$ and validation sets $\mathcal{D}_{V}$ are defined as

\begin{equation}
\mathcal{D}_k \equiv \bigcup_{i=1}^{k} D_i,
\end{equation}

Here, $\cup$ denotes the {set} union of the data, and the curriculum step is $k = \{1, 2, 3, 4, 5\}$. Each curriculum step in \texttt{FindCNN} lasts 6 epochs, minimizing a weighted \textit{BCEWithLogitsLoss} to lower the number of FPs, i.e. it is more relevant to correctly classify the noise class (negative class) than the injection class (positive class) {to avoid false alerts}. Thus, as in the original work, we weigh the injection class by 0.4. Regarding the optimizer, we employ AdaMax \cite{kingma2017adam} with a weight decay of $10^{-5}$ and the learning rate is $8\times 10^{-5}$. 

While \texttt{GWaveNet} has less trainable parameters than \texttt{FindCNN}, it is quite {a} complex model. To control the overfitting of the model we employ an early stopping algorithm where the curriculum step stops early if

\begin{equation}
|\mathcal{L}^{e}_{val} - \mathcal{L}^{e^{\text{best}}}_{val}| < \epsilon, \quad \text{ given } e - e^{\text{best}} \geq n, 
\label{eq:earlystopping}
\end{equation}
where $\mathcal{L}^{e}_{val}$ represents the validation loss of the current epoch $e$, $\mathcal{L}^{e^{\text{best}}}_{val}$ is the validation loss of the best epoch $e^{\text{best}}$, $\epsilon$ the tolerance and $n$ the patience. {Too small a tolerance gives way to overfitting and too high a tolerance does not let the NN reach its full potential, favouring underfitting, and the opposite goes for the patience.} After several experiments we set $\epsilon = 10^{-4}$ and $n=2${, which proved to be good trade-off values}. Moreover, for each curriculum step, we select the weights of the {best-performing} epoch{, i.e. the one} with {the} lowest {validation loss,} as initialization for the next step. {W}e {also} employ \textit{BCEWithLogitsLoss} {weighted at} $0.1$. After several experiments{,} {the} AdamW optimizer \cite{kingma2017adam} showed the best performance. Furthermore, we use an adaptive learning rate to allow the network to learn more slowly at more difficult steps: given an initial learning rate $lr_{0} = 2\times 10^5$, at each curriculum step $c \in [1, 5]$, we set $lr_{c} = lr_{0}/ c$.

\subsection{Modified architectures and quantization}
\label{sec:quantization}

In this work, we used the AMD Kria KV260 Vision AI Starter Kit \cite{amdKriaKV260}, and the Vitis AI software \cite{amdVitisAIDocs} for the deployment of the models. As Kria KV260 is designed for vision applications, it is only prepared to deal with 2-dimensional operations. Hence, to run the models in the chosen FPGA, we transformed the 1-dimensional \texttt{FindCNN} and \texttt{GWaveNet} to 2-dimensional versions: \texttt{\texttt{FindCNN}2D} and \texttt{GWaveNet2D}. Nonetheless, in a low-latency context, this implies reshaping within the FPGA, but such operation is not supported to run on the data processing unit (DPU){, needing to be run in the non-optimized CPU of the FPGA}. 

Further modifications were performed to maximize the number of layers that could run on the DPU. For \texttt{FindCNN}, we {added a $1\times1$ convolution at the end of the convolution blocks to cut the number of filters in half. On the other hand, \texttt{GWaveNet} required more changes: we moved from constant to the supported replicate padding, we changed {the} traditional $\tanh$ and $\sigma$ {activations} to linearized versions, we reduced the kernels size $k$ from 16 to 2, to allocate the model in memory{ and we lowered the number of filters in the sixth module to 32. This last step forced us to add a $1\times1$ convolutional layer after each \texttt{Wavenet} module to reduce the number of filters of each block's output to 32}. 
However, despite these changes, at the end of the day, {no versions of} \texttt{GWaveNet} {could be compiled {in the DPU IP} because of software limitations.}

Last but not least, due to the limited memory of FPGAs, it is necessary to quantize our models, transforming the model's weights from floating point to lower bit-widths to save memory and accelerate inference. In this work, we use the power-of-two quantization \cite{miyashita2016convolutionalneuralnetworksusing}, which is a robust logarithm-based quantization strategy implemented in Vitis AI \cite{xilinxvaiquantizer}.

\section{Results}
\label{sec:results}

\subsection{Performance}
In Section \ref{sec:dataset}{,} we mentioned that the input data has $155,648\,$data points and {three} channels, corresponding to Hanford, Livingston and Virgo, that we use to train, validate and test both models. In Fig. \ref{fig:loss_acc}, we present the losses (solid line) and accurac{ies} (dashed line) for \texttt{FindCNN} (top panel) and \texttt{GWaveNet} (bottom panel) for training (blue) and validation (yellow). {{For this figure, w}e use a decision threshold of $0.5$ { to calculate the accuracies,} i.e. the cutoff value used to {distinguish} the {classes} {is 0.5}.} The progressive curriculum learning step is shown by vertical dotted lines, where at each step we add samples further away from the merger. Furthermore, we mark the best-performing epoch {from each curriculum learning step} in red, i.e. the epoch with the best validation {loss}.

\begin{figure}[hbt!]
  \centering  
  \includegraphics[width=0.9\columnwidth]{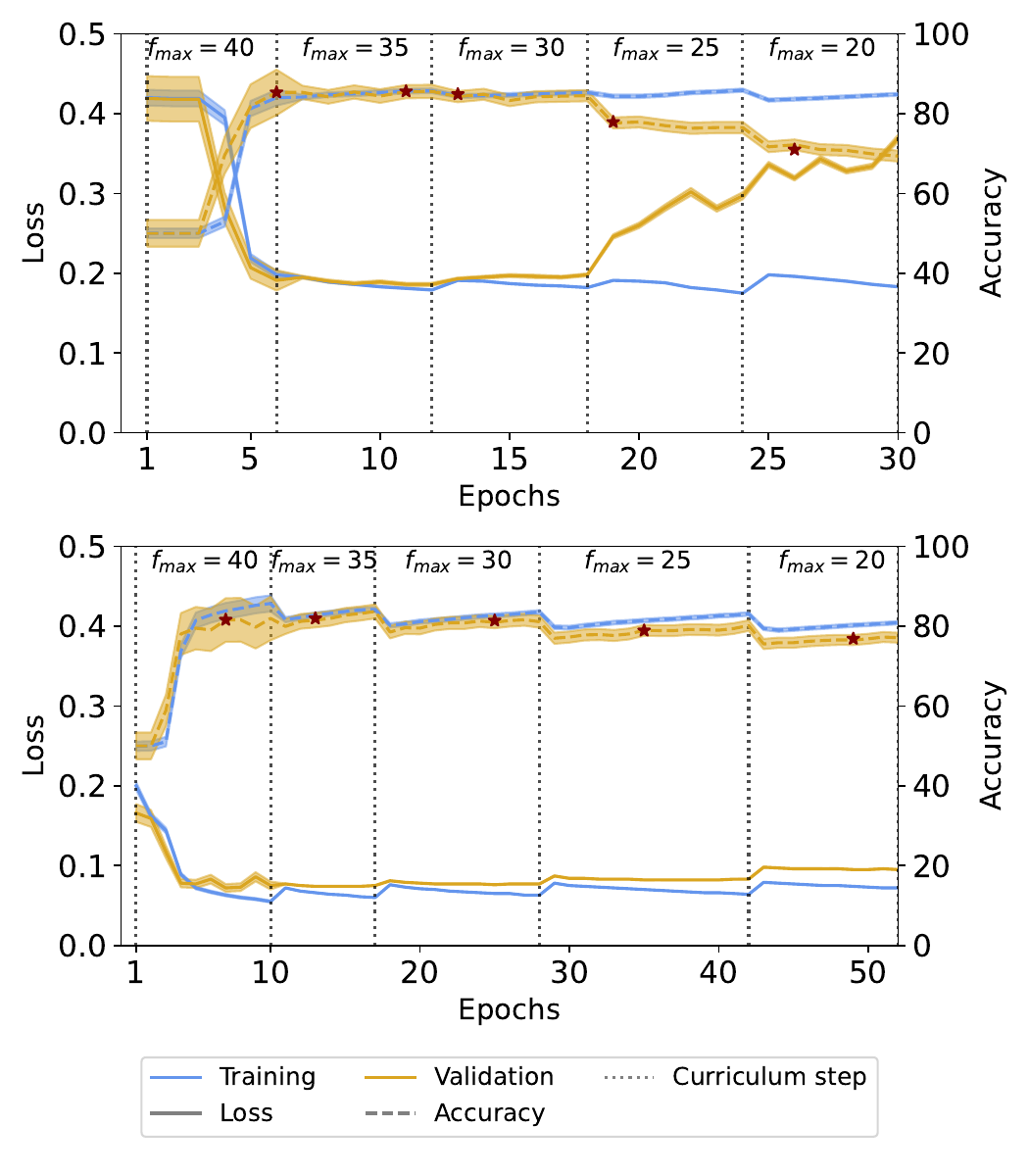}
  \caption{Training and validation of \texttt{FindCNN} (top) and \texttt{GWaveNet} (bottom).{ We present the average training (solid) and validation (yellow) loss{es} (solid line) and accurac{ies (dashed line) at the default threshold of 0.5}. The standard error is calculated at $3$ standard deviations $\sigma$ (shadowed region).} The curriculum learning step is shown by vertical dotted lines, and we mark the best-performing epoch {out of every step with a }red star, i.e. the epoch with the best validation accuracy up to a given tolerance (see Eq. \ref{eq:earlystopping}).} 
  \label{fig:loss_acc}
\end{figure}

For \texttt{FindCNN} (top panel) we can observe that at the beginning of the first curriculum step, both training and validation losses decrease rapidly, while the accuracy increases, {with} the best-performing epoch at the end of the step. A similar pattern is observed in the second step. However, by the third step {($f_{max} = 30\,$Hz)}, signs of overfitting emerge, with the best performance occurring at the beginning of the step. Overfitting becomes more pronounced in the fourth and fifth steps, evidenced by the divergence between training and validation loss and accuracy.

Such overfitting is not observed in \texttt{GWaveNet}, as training and validation are close together throughout the progressive curriculum learning steps. Additionally, the best-performing epoch typically occurs in the middle or at the end of each step. Note that, due to the early stopping method (see Section \ref{sec:learningstrategy}), the curriculum steps are shorter for easier{-to-learn} datasets, and longer for harder ones. Another notable detail is the expected increase in loss at the beginning of each curriculum learning step, followed by a decrease as learning {stabilises}, which is characteristic of curriculum learning strategies.

\begin{figure}[hbt!]
  \centering  
  \includegraphics[width=0.9\columnwidth]{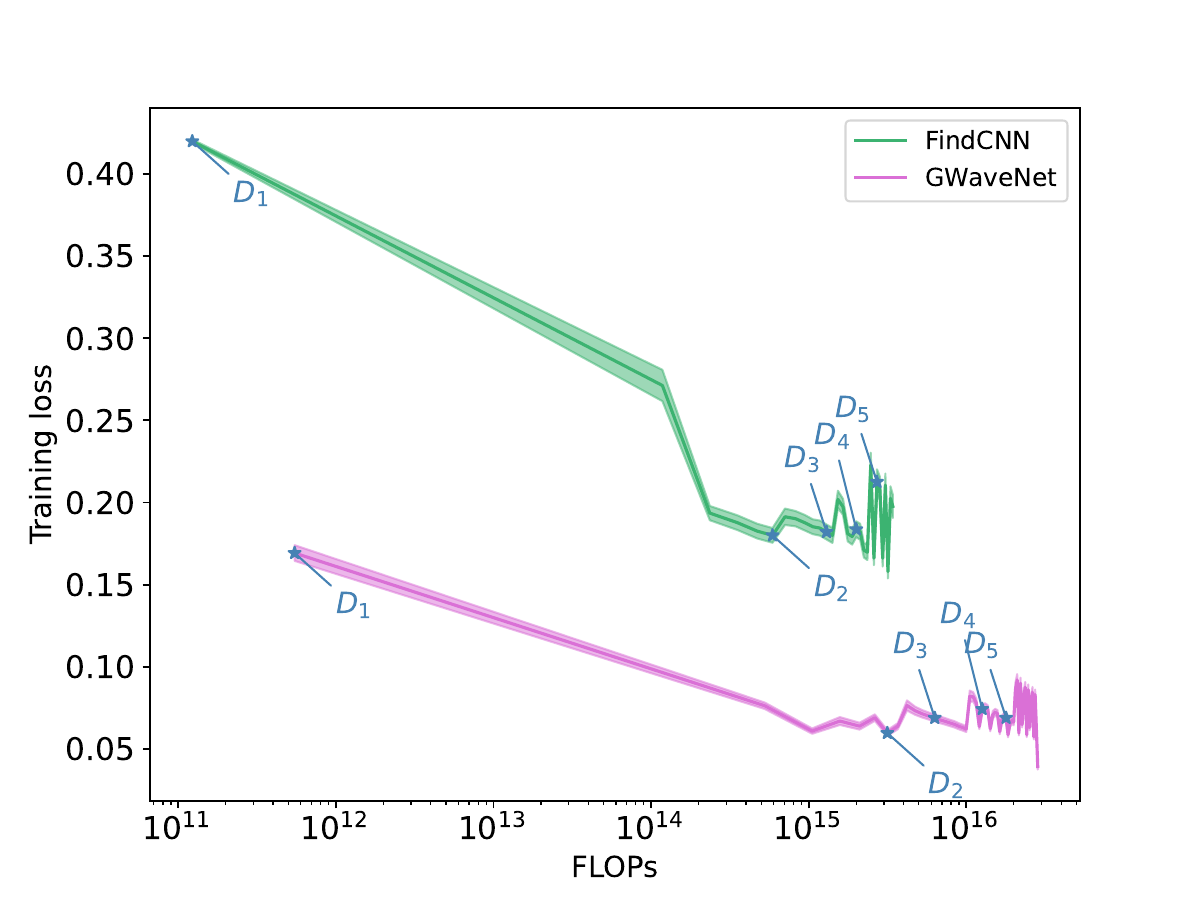}
  \caption{ Training losses over the number of floating point operations (FLOPs) in logarithmic scale for \texttt{FindCNN} (green) and \texttt{GWaveNet} (pink). We present the average loss per epoch for the cumulative amount of the number of FLOPs and $3$ standard deviations $\sigma$. We mark with a blue star the beginning of each curriculum step.} 
  \label{fig:loss_flops}
\end{figure}

Based on {these observations}, we can conclude that \texttt{GWaveNet} outperforms \texttt{FindCNN} with fewer trainable parameters{{, potentially implying lower computational cost}}. In Fig. \ref{fig:loss_flops}, we present the training loss as a function of the number of floating point operations (FLOPs) for \texttt{FindCNN} (green) and \texttt{GWaveNet} (pink), {and also mark in blue the begininning of each curriculum step. We can observe that for $D_{1}$ and $D_{2}$ curriculum steps the learning is smooth as the loss is minimized. However, for the hardest curriculum learning steps the loss becomes quite unstable, particularly $D_{5}$ for both models.} From this plot, it is evident that \texttt{GWaveNet} performs more FLOPs than \texttt{FindCNN}, {in part because it trains for more epochs, {but also since the beginning of the training due to its architecture}}. {Still}, \texttt{GWaveNet} achieves {a better} loss reduction with a comparable number of FLOPs.

\begin{figure}[hbt!]
  \centering  
  \includegraphics[width=1\columnwidth]{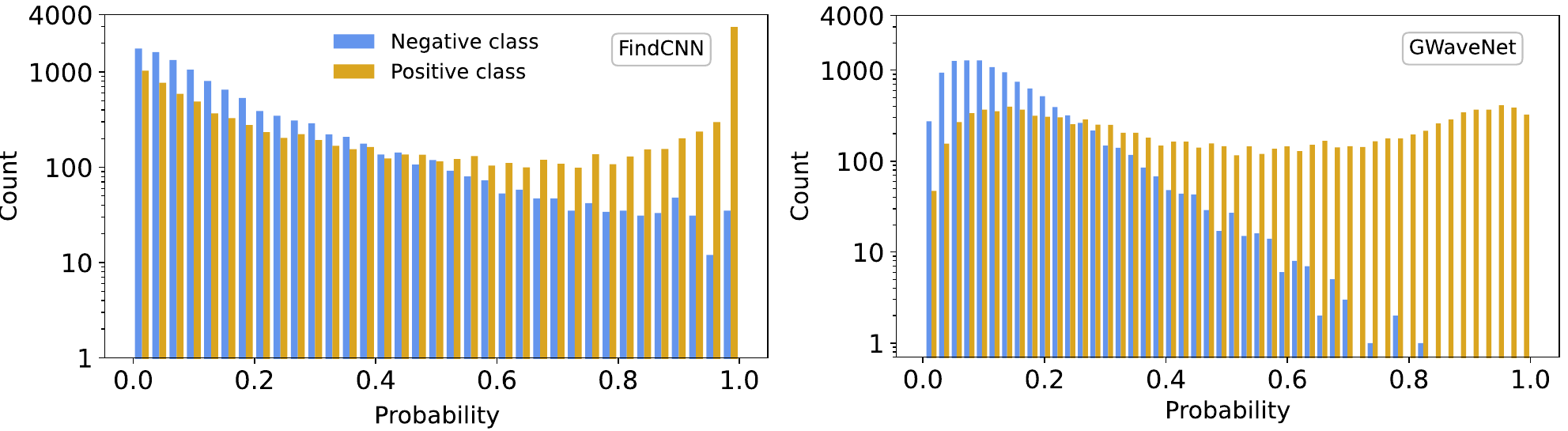}
  \caption{ {Distribution of the {score for if} a GW signal {is} present, as determined by each model. We show  the noise class in yellow, and noise + signal class in blue.} } 
  \label{fig:class_probability}
\end{figure}

While accuracy and loss functions are relevant magnitudes to measure performance, we are interested in minimizing the number of FP{s}. To assess the separability of both classes, in Fig. \ref{fig:class_probability} we show the noise (negative class{, in blue}) and the noise + GW signal (positive class{, in yellow}) of the testing set as a function of the {score of if it} contain{s} a GW signal. We can observe that \texttt{FindCNN} is more decisive than \texttt{GWaveNet}, as it usually provides extreme {scores}, either 0 or 1. Nonetheless, this decisiveness is a symptom of being a ``yes-classifier'', as many noise samples are classified with high {scores}. On the other hand, \texttt{GWaveNet} is more conservative and is less keen on providing extreme {scores \sout{values}}{, a desired behaviour} {for this particular application}. Furthermore, for \texttt{GWaveNet} there is a {notable separation} between both classes for {scores} $>0.81$. Note that both models use a weighted loss function: \texttt{FindCNN} is weighted by 0.4, while \texttt{GWaveNet} is weighted by 0.1.

\begin{figure}[hbt!]
  \centering  
  \includegraphics[width=1\columnwidth]{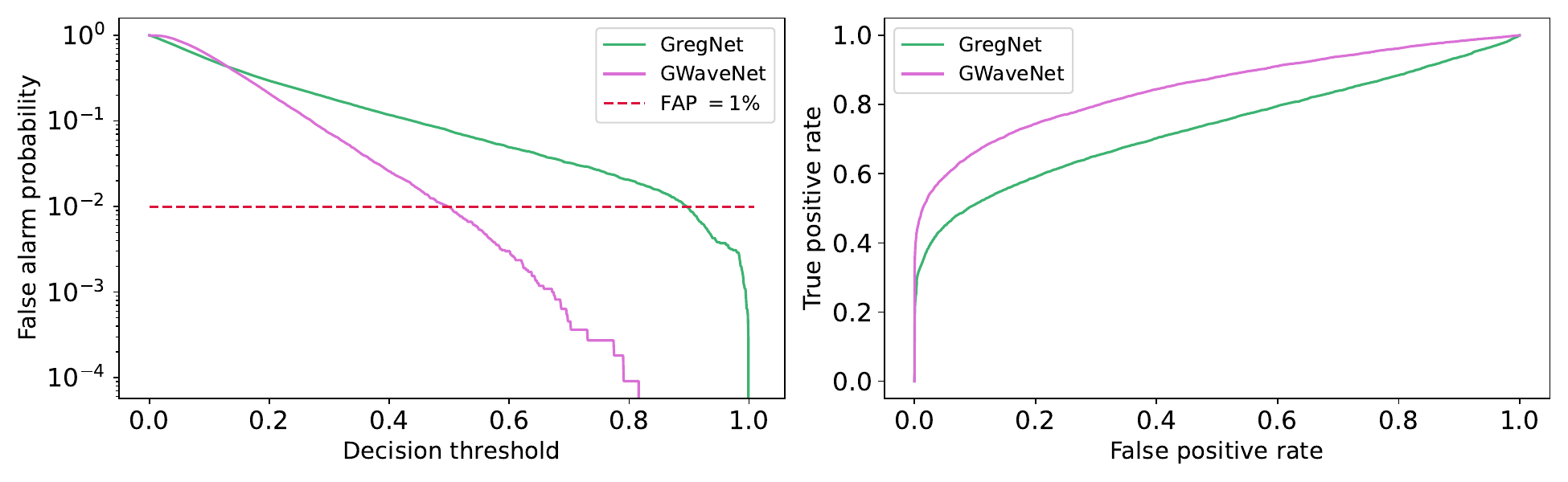}
  \caption{\textit{({Left})} False alarm probability (FAP) -- or false positive rate -- in logarithmic scale as a function of the decision threshold for \texttt{FindCNN} (green) and GWaveNet (pink), marking the threshold FAP$= 1\%$ in red. \textit{({Right})} Receiver operator characteristic (ROC) curve of \texttt{FindCNN} (green) and \texttt{GWaveNet} (pink).} 
  \label{fig:fap_comparison_roc_curve}
\end{figure}

In a GW experiment{,} it is crucial assess the number of FP. A common metric in GW is the false alarm probability (FAP), also known as false positive rate. In {the {left-hand} panel of} Fig. \ref{fig:fap_comparison_roc_curve}} we plot FAP as a function of the decision threshold for \texttt{FindCNN} (green) and \texttt{GWaveNet} (pink). We mark the target FAP $=1\%$ in red, as defined in \cite{Baltus:2022pep}. We can observe that \texttt{GWaveNet} achieves FAP $=1\%$ for a decision threshold of 0.46, while \texttt{FindCNN} needs a more aggressive decision threshold of 0.86. Furthermore, \texttt{GWaveNet} allows to set FAP $\leq10^{-4}$. 

In ML  it is standard to use the receiver operator characteristic (ROC) curve, i.e. true positive rate as a function of false positive rate for different decision threshold steps. In {the right-hand panel of} Fig. \ref{fig:fap_comparison_roc_curve} we show the ROC curve of \texttt{FindCNN} (green) and \texttt{GWaveNet} (pink), and we can observe how \texttt{GWaveNet} achieves a larger area under the curve, outperforming \texttt{FindCNN}.

\begin{figure}[hbt!]
  \centering  
  \includegraphics[width=1\columnwidth]{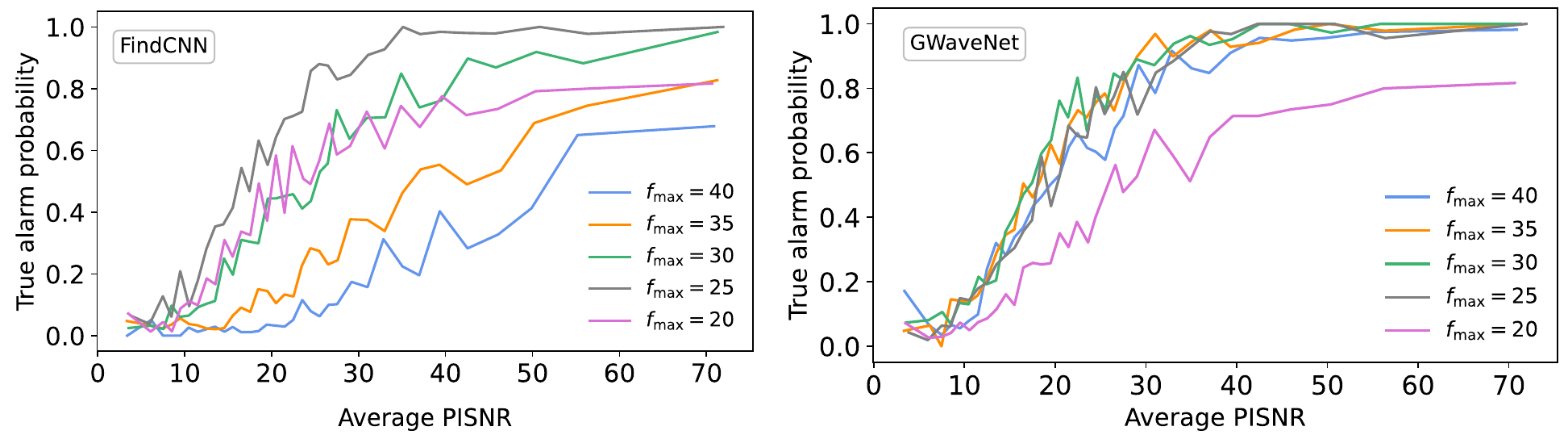}
  \caption{{True alarm probability (TAP)-or true positive rate-as a function of PISNR for {\texttt{FindCNN} ({left}) and \texttt{GWaveNet} ({right})} for different curriculum steps-stepping on average maximum frequency $f_{\text{max}}$-at} 
false alarm probability (FAP) of $1\%$.} 
  \label{fig:probability_pisnr}
\end{figure}

One of the the main limitations of \texttt{FindCNN} (see Section \ref{sec:SOTA}), is that the network forgets the previous curriculum learning steps. To compare the ``memory" of \texttt{FindCNN}, we reproduce Fig. 5 of \cite{Baltus:2022pep} in Fig. \ref{fig:probability_pisnr}. Here, we represent the true alarm probability (TAP), also known as the true positive rate, as a function of the average PISNR for the final models after the progressive curriculum learning strategy. In {the {left-hand} panel of} Fig. \ref{fig:probability_pisnr}, we can see that the final \texttt{FindCNN} has a {TAP $> 0.5$ for signals with PISNR $> 30$} and $f_{max} = {20, 25, 30}\,$Hz, the most difficult samples. As expected, TAP decreases for lower PISNR, which corresponds to the most quiet signals. Nonetheless, {its performance drops $\sim 20$ percentage points (p.p.) for $f_{max} = {35, 40}\,$Hz at $\approx 25$ PISNR, due to catastrophic forgetting}. In a realistic GW search, this would be {unreliable}, as the algorithm would not alert about the signals closer to the merger.
On the other hand, we can observe how the final \texttt{GWaveNet} can ``remember" easier data sets while maintaining a comparable performance to \texttt{FindCNN} for the most difficult data set $f_{max} =20\,$Hz.

As we present in Table \ref{tab:accuracies-fap0.01}, \texttt{GWaveNet} outperforms \texttt{FindCNN} by $10\,$p.p., but these models are unable to run on an FPGA due to memory limitations. {I}n Table, \ref{tab:accuracies-fap0.01} we also show the accuracies of the modified models before and after quantization. {See Section \ref{sec:quantization} for further details on memory limitations and quantization strategy.} It is relevant to note that
\texttt{{FindCNN}2D} and \texttt{GWaveNet2D} outperform their 1-dimensional versions by $\sim 3\,$p.p., which could imply that Pytorch \cite{paszke2019pytorch} is better optimized for vision applications. This could {also} result in better numerical stability, leading to higher accuracies, but further investigation is needed. On the other hand, the modified versions perform worse than the original ones, due to a simplification in certain operations{, such as lowering the size of the dense layers for \texttt{FindCNN}, and changing the padding mode, changing the activation functions to linear equivalents and also lowering the size of the dense layers for \texttt{GWaveNet}}. This is particularly notorious for \texttt{GWaveNet2DModified}, which has $\sim 5\,$p.p. less accuracy than original \texttt{GWaveNet}, but its performance improves $\sim 1\,$p.p. after quantization{, likely due to the smoothing of the {possibly overfit} learnt manifold}. Nonetheless, the worst-performing \texttt{GWaveNet} still outperforms the best-performing \texttt{FindCNN} by $\sim 2\,$p.p.

\begin{table}[hbt!]
\centering
\begin{tabular}{|c|c|c|}
\hline
  \bf{Model}             & \begin{tabular}[c]{@{}c@{}}\bf{Before}\\\bf{quantization}\end{tabular}  & \begin{tabular}[c]{@{}c@{}}\bf{After}\\ \bf{quantization}\end{tabular} \\ \hline
{FindCNN}            & 66.81\%                             & -                                  \\
{FindCNN}2D          & \textbf{69.50}\%                             & 69.37\%                                   \\
\begin{tabular}[c]{@{}c@{}}{FindCNN}2DModified\end{tabular}   &  66.70\%                                   & 65.07\%                                   \\
GWaveNet           & 76.22\%                             & -                                  \\
GWaveNet2D         &  79.42\%                                  &  77.28\%                                 \\
\begin{tabular}[c]{@{}c@{}}GWaveNet2DModified\end{tabular}  & \textbf{71.72\%}                                  & 72.42\%                                   \\ \hline
\end{tabular}
\caption{Test accuracies at {FAP} set at 1\% before and after quantization. All models were tested on 22000 samples, except for the \texttt{GWaveNet2D} and \texttt{GWaveNet2DModified} post-quantization models, which were tested on 5000 samples, due to timeout constraints. We highlight in bold the best FindCNN model and the worst GWaveNet model. }.\label{tab:accuracies-fap0.01}
\end{table}

\begin{table}[hbt!]
\resizebox{\textwidth}{!}{ 
\begin{tabular}{|c|ccc|ccc|ccc|}
\hline
\multirow{2}{*}{\textbf{Model}}                                & \multicolumn{3}{c|}{\textbf{Time (ms)}}     & \multicolumn{3}{c|}{\textbf{Energy consumption (mJ)}} & \multicolumn{3}{c|}{\textbf{Economic cost (\$/year)}} \\ \cline{2-10} 
                                                               & CPU           & GPU          & FPGA         & CPU           & GPU            & FPGA          & CPU           & GPU           & FPGA         \\ \hline
FindCNN                                                        & 27.53 ± 3.69  & 21.75 ± 1.89 & -            & 107.24 ± 4.26 & 617.12 ± 37.11 & -             & 28.83         & 210.01        & -            \\ \hline
FindCNN2D                                                      & 29.58 ± 0.10  & 20.90 ± 1.56 & 84.54 ± 0.10 & 177.93 ± 1.60 & 606.74 ± 32.06 & 468.59 ± 2.40 & 44.52         & 214.87        & 41.03        \\ \hline
\begin{tabular}[c]{@{}c@{}}FindCNN-\\ 2DModified\end{tabular}  & 29.54 ± 0.11  & 24.87 ± 1.65 & 84.28 ± 0.10 & 178.35 ± 1.34 & 694.59 ± 28.46 & 471.51 ± 2.96 & 44.69         & 206.72        & 41.41        \\ \hline
GWaveNet                                                       & 154.75 ± 0.38 & 33.60 ± 2.76 & -            & 583.66 ± 5.56 & 983.83 ± 60.52 & -             & 27.92         & 216.72        & -            \\ \hline
GWaveNet2D                                                     & 85.28 ± 2.87  & 29.41 ± 1.77 & -            & 568.92 ± 4.95 & 858.11 ± 36.83 & -             & 49.38         & 215.96        & -            \\ \hline
\begin{tabular}[c]{@{}c@{}}GWaveNet-\\ 2DModified\end{tabular} & 65.54 ± 3.67  & 26.02 ± 3.25 & -            & 240.76 ± 5.85 & 752.49 ± 57.90 & -             & 27.19         & 214.05        & -            \\ \hline
\end{tabular}
}
\caption{Average time (ms) and energy consumption (mJ) to predict a single sample and economic cost of running each model in a year ($\$$/year) for different devices (see details in Section \ref{sec:hardware}) {at 3 standard deviations given 10 experiments. {Note that only FindCNN2D and FindCNN2DModified was compatible with our quantization framework.}}}\label{tab:cost}
\end{table}

\subsection{Time, energy and economic cost}

To compare the models' efficiencies, we measured the time taken to test 22,000 samples on both the CPU and GPU and 1,000 samples on the FPGA. The results of time (ms), energy consumption (mJ), and cost ($\$$/year) to run each model for CPU, GPU and FPGA are presented in Table \ref{tab:cost}. As mentioned in Section \ref{sec:quantization}, we were unable to compile \texttt{GWaveNet} in the FPGA due to software limitations, potentially due to the lack of support of the padding needed for causal convolutions, but further investigation is required.

Regarding the average inference time, we can observe how all \texttt{FindCNN} models take $\sim 25\,$ms to predict a sample in CPU and GPU. However, the time in an FPGA is 4 times longer. Further exploration revealed that the bottleneck was caused by the batch normalization layer in Vitis AI \cite{amdVitisAIPyTorchOperators}. Ignoring this layer decreased the FPGA times and energy consumption in half. Nonetheless, the lack of speed of the FPGA could be due to {our particular low-end FPGA}, which was not designed for processing high-dimensional time series data.

For \texttt{GWaveNet}{,} the inference time on GPU is slightly higher than \texttt{FindCNN}, taking 1.5 times longer. Notably, on the CPU, original \texttt{GWaveNet} takes $\sim 2$ times more than the 2-dimensional versions, {making} \texttt{GWaveNet2DModified} the fastest, due to its simplified nature. It is also interesting to note that on the GPU the original models are slightly slower than their 2-dimensional equivalents. As we hypothesized in the previous section, this could be attributed to PyTorch's enhanced optimization for vision applications.
However, it is worth noting that, given the input duration of $300\,$s, the time consumption of all the models is remarkably low.

Regarding the energy consumption needed for inference (see Table \ref{tab:cost}), when comparing the same model across various devices, we can see that while the GPU is one of the fastest, it is also the one that consumes the most energy. As before, if we ignore the batch normalization layers of \texttt{FindCNN} models on the FPGA, the energy consumption is halved, being comparable to the CPU consumption. Thus, we believe that a model fully optimized for {an} FPGA would have a better energy performance, and this will be explored in future works. 

In a realistic situation, these models would run through the whole observing run. In Table \ref{tab:cost}, we consider the economic cost of running the models for a year, taking the average industry electricity price in the USA, as most GW clusters are there. In 2022, the average electricity cost was $0.0845 \,$\$/kWh \cite{statista_energy_prices_us}. As expected, the GPU is the most expensive. {On the other hand, the cost {of running on an FPGA is} $\approx 7\%$ cheaper than running on a CPU.} Note that we only take into account the electricity prices, but adding the upfront and upkeep prices of the devices, the FPGA is by far the least expensive. {Assuming the price at launch date over the number of lifetime years, under heavy use, and the prices for running \texttt{FindCNN2D} for a year, running on a CPU ($\approx 495\$$) or GPU ($\approx 465\$$) have similar expenses, whereas running on an FPGA ($\approx 65\$$) costs over seven times less.}

Lastly, running \texttt{FindCNN} and \texttt{GWaveNet} incurs comparable economic costs. This highlights \texttt{GWaveNet}'s efficiency, achieving better performance without significantly increasing computational resources, making it a more cost-effective model for ML-based BNS search algorithms.

\section{Conclusions}
\label{sec:conclusions}

In this work{, we }develop{ed} a novel ML-based GW detection algorithm for the early inspiral of {BNS} mergers, \texttt{GWaveNet}, inspired by the audio generating \texttt{WaveNet} \cite{oord2016wavenet}. We compare its performance to the current state-of-the-art, \texttt{FindCNN} \cite{Baltus:2022pep}, reproducing this work in simulated detector noise from the third observing run. Moreover, we compare their time, energy consumption and cost in CPU, GPU and FPGA. 

As in \cite{Baltus:2022pep}, we implement a progressive curriculum learning strategy, where easier (more difficult) samples have higher (lower) maximum frequency $f_{max}$, and are closer (further) to the merger, meaning that they will be detected later (earlier). Both \texttt{FindCNN} and \texttt{GWaveNet} are binary classifiers, that differentiate between noise (negative class) and noise + signal (positive class) solely utilizing the inspiral of {BNS} mergers. 

\texttt{GWaveNet} has less trainable parameters than \texttt{FindCNN}, but it is deeper and more complex, implementing key ideas to improve its performance at a limited computational cost. Furthermore, we implement a more flexible training strategy, accommodating \texttt{GWaveNet} to the difficulty of each curriculum step. In terms of accuracy at FAP=1\%, \texttt{FindCNN} achieves 66.81\%, while \texttt{GWaveNet} achieves 76.22\%, outperforming \texttt{FindCNN} by $\sim 10\,$p.p. In terms of inference time, \texttt{FindCNN} is $\sim 2\,$ms faster on GPU, but we believe \texttt{GWaveNet} is the preferred model due to its enhanced performance.

Regarding the FPGA, while we were able to test the performance of \texttt{FindCNN}, this was not possible with \texttt{GWaveNet} due to software limitations. 
In terms of costs, the FPGA is the most sustainable option. Although the GPU is the fastest device, it also has the highest energy consumption.

The present work demonstrated high performance in terms of accuracy at a limited computational cost. In future works, we will study more realistic scenarios, moving to real detector data. Furthermore, other FPGA software and/or hardware, better fitting for our application{, could also be explored}.

\section*{Acknowledgments}

The authors thank M.\ Hester and R.\ Aaij for the fruitful and inspiring discussions during this study. This project was supported by Nikhef Laboratory, and the authors extend their gratitude to the Nikhef computing group.  This publication is part of the project Computing for Virgo with file number 2023.033 which is financed by the Dutch Research Council (NWO). M.L.\ is supported by the research program of the Netherlands Organisation for Scientific Research (NWO).  MvdS acknowledges support from the European Union, AHEAD~2020 (grant number 871158).  This publication is part of the project Cortex with project number 00686766 of the research programme NWA which is (partly) financed by the Dutch Research Council (NWO).  This material is based upon work supported by NSF’s LIGO Laboratory which is a major facility fully funded by the National Science Foundation. 

\printbibliography[heading=bibintoc]

@article{Buonanno:2002fy,
    author = "Alessandra Buonanno and Yanbei Chen and Michele Vallisneri",
    title = "{Detecting gravitational waves from precessing binaries of spinning compact objects: Adiabatic limit}",
    eprint = "gr-qc/0211087",
    archivePrefix = "arXiv",
    doi = "10.1103/PhysRevD.67.104025",
    journal = "Phys. Rev. D",
    volume = "67",
    pages = "104025",
    year = "2003",
    note = "[Erratum: Phys.Rev.D 74, 029904 (2006)]"
}

@article{Biwer_2019,
   title={PyCBC Inference: A Python-based Parameter Estimation Toolkit for Compact Binary Coalescence Signals},
   volume={131},
   ISSN={1538-3873},
   DOI={10.1088/1538-3873/aaef0b},
   number={996},
   journal={Publications of the Astronomical Society of the Pacific},
   publisher={IOP Publishing},
   author={C. M. Biwer and Collin D. Capano and Soumi De and Miriam Cabero and Duncan A. Brown and Alexander H. Nitz and V. Raymond},
   year={2019},
   month=jan, pages={024503} }

@misc{amdKriaKV260,
  title={{AMD Kria KV260 Vision Starter Kit}},
  author={{Advanced Micro Devices, Inc.}},
  year={Accessed 27 March 2024},
  url={https://www.amd.com/en/products/system-on-modules/kria/k26/kv260-vision-starter-kit.html},
}

@article{KAGRA:2023pio,
    author = "Abbott, R. and others",
    collaboration = "KAGRA, VIRGO, LIGO Scientific",
    title = "{Open Data from the Third Observing Run of LIGO, Virgo, KAGRA, and GEO}",
    eprint = "2302.03676",
    archivePrefix = "arXiv",
    primaryClass = "gr-qc",
    reportNumber = "LIGO-P2200316",
    doi = "10.3847/1538-4365/acdc9f",
    journal = "Astrophys. J. Suppl.",
    volume = "267",
    number = "2",
    pages = "29",
    year = "2023"
}

@article{Tsukada:2023edh,
    author = "L. Tsukada and others",
    title = "{Improved ranking statistics of the GstLAL inspiral search for compact binary coalescences}",
    eprint = "2305.06286",
    archivePrefix = "arXiv",
    primaryClass = "astro-ph.IM",
    doi = "10.1103/PhysRevD.108.043004",
    journal = "Phys. Rev. D",
    volume = "108",
    number = "4",
    pages = "043004",
    year = "2023"
}

@article{Magee:2021xdx,
    author = "R. Magee and others",
    title = "{First demonstration of early warning gravitational wave alerts}",
    eprint = "2102.04555",
    archivePrefix = "arXiv",
    primaryClass = "astro-ph.HE",
    doi = "10.3847/2041-8213/abed54",
    journal = "Astrophys. J. Lett.",
    volume = "910",
    number = "2",
    pages = "L21",
    year = "2021"
}

@article{Yu:2021vvm,
    author = "Hang Yu and Rana X. Adhikari and Ryan Magee and Surabhi Sachdev and Yanbei Chen",
    title = "{Early warning of coalescing neutron-star and neutron-star-black-hole binaries from the nonstationary noise background using neural networks}",
    eprint = "2104.09438",
    archivePrefix = "arXiv",
    primaryClass = "gr-qc",
    doi = "10.1103/PhysRevD.104.062004",
    journal = "Phys. Rev. D",
    volume = "104",
    number = "6",
    pages = "062004",
    year = "2021"
}

@article{LIGOScientific:2020ibl,
    author = "Abbott, R. and others",
    collaboration = "LIGO Scientific, Virgo",
    title = "{GWTC-2: Compact Binary Coalescences Observed by LIGO and Virgo During the First Half of the Third Observing Run}",
    eprint = "2010.14527",
    archivePrefix = "arXiv",
    primaryClass = "gr-qc",
    reportNumber = "P2000061",
    doi = "10.1103/PhysRevX.11.021053",
    journal = "Phys. Rev. X",
    volume = "11",
    pages = "021053",
    year = "2021"
}

@misc{amdVitisAIDocs,
  title={{AMD Vitis AI Documentation}},
  author={{Advanced Micro Devices, Inc.}},
  year={Accessed 02 April 2024},
  url={https://docs.amd.com/r/3.0-English/ug1414-vitis-ai},
}

@article{paszke2019pytorch,
      title={PyTorch: An Imperative Style, High-Performance Deep Learning Library}, 
      author={ A. Paszke and others},
      year={2019},
      pages={1912.01703},
      archivePrefix={arXiv},
      primaryClass={cs.LG}, 
journal = {arXiv e-prints},
}

@article{article,
author = {Ali, Asad and Ahmad, Salman and Nawaz, M. and Ullah, Saleem and Aqeel, Muhammad},
year = {2015},
month = {December},
pages = {645},
title = {Bayesian Inference on Gravitational Waves},
volume = {11},
journal = {Pakistan Journal of Statistics and Operation Research},
doi = {10.18187/pjsor.v11i4.1053}
}

@article{PhysRevLett.116.061102,
  title = {Observation of Gravitational Waves from a Binary Black Hole Merger},
  author = {Abbott, B. P. and others},
  collaboration = {LIGO Scientific Collaboration and Virgo Collaboration},
  journal = {Phys. Rev. Lett.},
  volume = {116},
  issue = {6},
  pages = {061102},
  numpages = {16},
  year = {2016},
  month = {Feb},
  publisher = {American Physical Society},
  doi = {10.1103/PhysRevLett.116.061102},
}

@misc{amdVitisAIPyTorchOperators,
  title={{AMD Vitis AI Documentation - Operators Supported by PyTorch}},
  author={{Advanced Micro Devices, Inc.}},
  year={Accessed 2024},
  url={https://docs.amd.com/r/3.0-English/ug1414-vitis-ai/Operators-Supported-by-PyTorch},
}

@article{Que_2024,
   title={LL-GNN: Low Latency Graph Neural Networks on FPGAs for High Energy Physics},
   volume={23},
   ISSN={1558-3465},
   DOI={10.1145/3640464},
   number={2},
   journal={ACM Transactions on Embedded Computing Systems},
   publisher={Association for Computing Machinery (ACM)},
   author={Zhiqiang Que and Hongxiang Fan and Marcus Loo and He Li and Michaela Blott and Maurizio Pierini and Alexander Tapper and Wayne Luk},
   year={2024},
   month=mar, pages={1–28} }

@article{khoda2022ultralow,
   author = "Khoda, Elham E. and others",
    title = "{Ultra-low latency recurrent neural network inference on FPGAs for physics applications with hls4ml}",
    eprint = "2207.00559",
    archivePrefix = "arXiv",
    primaryClass = "cs.LG",
    doi = "10.1088/2632-2153/acc0d7",
    journal = "Mach. Learn. Sci. Tech.",
    volume = "4",
    number = "2",
    pages = "025004",
    year = "2023"
}

@inproceedings{fpga-nns-paper,
 author={Ruoyu Sang and Qiang Liu and Qijun Zhang},
  booktitle={2016 IEEE MTT-S International Conference on Numerical Electromagnetic and Multiphysics Modeling and Optimization (NEMO)}, 
  year={2016},
  volume={},
  number={},
  pages={1-2},
  doi={10.1109/NEMO.2016.7561676}, 
title={}, }

@inproceedings{10.5555/3104322.3104425,
author = {Nair, V. and Hinton, Geoﬀrey E.},
year = {2010},
isbn = {9781605589077},
publisher = {Omnipress},
address = {Madison, WI, USA},
booktitle = {Proceedings of the 27th International Conference on International Conference on Machine Learning},
pages = {807–814},
numpages = {8},
location = {Haifa, Israel},
series = {ICML'10},
title = {},
}

@article{kingma2017adam,
      title={Adam: A Method for Stochastic Optimization}, 
      author={Diederik P. Kingma and Jimmy Ba},
      year={2017},
      pages={1412.6980},
      archivePrefix={arXiv},
      primaryClass={cs.LG},
    journal = {arXiv e-prints},
}

@article{ioffe2015batch,
      title={Batch Normalization: Accelerating Deep Network Training by Reducing Internal Covariate Shift}, 
      author={Sergey Ioffe and Christian Szegedy},
      year={2015},
      pages={1502.03167},
      archivePrefix={arXiv},
      primaryClass={cs.LG},
      journal = {arXiv e-prints},
}

@misc{pytorchMaxPool1d,
  title={{PyTorch Documentation - torch.nn.MaxPool1d}},
  author={{PyTorch}},
  year={Accessed 22 April 2024},
  url={https://pytorch.org/docs/stable/generated/torch.nn.MaxPool1d.html},
}

@article{Meijer:2023yhn,
    author = "Quirijn Meijer and Melissa Lopez and Daichi Tsuna and Sarah Caudill",
    title = "{Gravitational-wave searches for cosmic string cusps in Einstein Telescope data using deep learning}",
    eprint = "2308.12323",
    archivePrefix = "arXiv",
    primaryClass = "astro-ph.IM",
    doi = "10.1103/PhysRevD.109.022006",
    journal = "Phys. Rev. D",
    volume = "109",
    number = "2",
    pages = "022006",
    year = "2024"
}

@misc{pytorchBCEWithLogitsLoss,
  title={{PyTorch Documentation - torch.nn.BCEWithLogitsLoss}},
  author={{PyTorch}},
  year={Accessed 24 April 2024},
  url={https://pytorch.org/docs/stable/generated/torch.nn.BCEWithLogitsLoss.html},
}

@article{Wei_2021,
   title={Deep learning ensemble for real-time gravitational wave detection of spinning binary black hole mergers},
   volume={812},
   ISSN={0370-2693},
   DOI={10.1016/j.physletb.2020.136029},
   journal={Physics Letters B},
   publisher={Elsevier BV},
   author={Wei Wei and Asad Khan and E. A. Huerta and Xiaobo Huang and Minyang Tian},
   year={2021},
   month=jan, pages={136029} }

@misc{xilinxvaiquantizer,
  title={Vitis-AI Quantization Configuration},
  author={{Xilinx, Inc.}},
  year={Accessed 07 May 2024},
  url={https://github.com/Xilinx/Vitis-AI/blob/3.0/src/vai_quantizer/vai_q_pytorch/doc/Quant_Config.md},
}

@article{Akhshi2020ATA,
    author = "A. Akhshi and H. Alimohammadi and S. Baghram and S. Rahvar and M. R. Rahimi Tabar and H. Arfaei",
    title = "{A template-free approach for waveform extraction of gravitational wave events}",
    pages = "2005.11352",
    archivePrefix = "arXiv",
    primaryClass = "astro-ph.IM",
    reportNumber = "Scientific Reports 11, 20507 (2021)",
    journal="arXiv pre-print",
    month = "5",
    year = "2020"
}

@misc{nvidia_gtx_1080,
  author = {{NVIDIA}},
  title = {GeForce GTX 1080 Specifications},
  year = {2024},
  howpublished = {\url{https://www.nvidia.com/en-gb/geforce/graphics-cards/geforce-gtx-1080/specifications/}},
  note = {Accessed 09 June 2024}
}

@article{Dooney:2024pvt,
    author = "Dooney, Tom and Curier, R. Lyana and Tan, Daniel Stanley and Lopez, Melissa and Van Den Broeck, Chris and Bromuri, Stefano",
    title = "{One flexible model for multiclass gravitational wave signal and glitch generation}",
    eprint = "2401.16356",
    archivePrefix = "arXiv",
    primaryClass = "physics.ins-det",
    doi = "10.1103/PhysRevD.110.022004",
    journal = "Phys. Rev. D",
    volume = "110",
    number = "2",
    pages = "022004",
    year = "2024"
}

@misc{microway_tesla_v100,
  author = {{Microway, Inc.}},
  title = {NVIDIA Tesla V100 Price Analysis},
  year = {2018},
  howpublished = {\url{https://www.microway.com/hpc-tech-tips/nvidia-tesla-v100-price-analysis/}},
  note = {Accessed 09 June 2024}
}

@misc{mitnews_carbonfootprint,
  author = {MIT News Office},
  title = {How can we reduce the carbon footprint of global computing?},
  year = {2022},
  howpublished = {\url{https://news.mit.edu/2022/how-can-we-reduce-carbon-footprint-global-computing-0428}},
  note = {Accessed 10 June 2024}
}

@misc{techpowerup_epyc7551p,
  author = {TechPowerUp},
  title = {AMD EPYC 7551P},
  year = {2024},
  howpublished = {\url{https://www.techpowerup.com/cpu-specs/epyc-7551p.c1929}},
  note = {Accessed 23 June 2024}
}

@misc{statista_energy_prices_us,
  author = {{Statista}},
  title = {Energy Prices in the U.S.},
  year = {2024},
  howpublished = {\url{https://www.statista.com/topics/6337/energy-prices-in-the-us/\#topicOverview}},
  note = {Accessed 23 June 2024}
}

@article{VIRGO:2014yos,
    author = "F. Acernese and others",
    collaboration = "VIRGO",
    title = "{Advanced Virgo: a second-generation interferometric gravitational wave detector}",
    eprint = "1408.3978",
    archivePrefix = "arXiv",
    primaryClass = "gr-qc",
    doi = "10.1088/0264-9381/32/2/024001",
    journal = "Class. Quant. Grav.",
    volume = "32",
    number = "2",
    pages = "024001",
    year = "2015"
}

@article{LIGOScientific:2014pky,
    author = "J. Aasi and others",
    collaboration = "LIGO Scientific",
    title = "{Advanced LIGO}",
    eprint = "1411.4547",
    archivePrefix = "arXiv",
    primaryClass = "gr-qc",
    doi = "10.1088/0264-9381/32/7/074001",
    journal = "Class. Quant. Grav.",
    volume = "32",
    pages = "074001",
    year = "2015"
}

@article{Reitze:2019iox,
    author = "Reitze, D. and others",
    title = "{Cosmic Explorer: The U.S. Contribution to Gravitational-Wave Astronomy beyond LIGO}",
    eprint = "1907.04833",
    archivePrefix = "arXiv",
    primaryClass = "astro-ph.IM",
    reportNumber = "LIGO-P1900316",
    journal = "Bull. Am. Astron. Soc.",
    volume = "51",
    number = "7",
    pages = "035",
    year = "2019"
}

@article{Punturo:2010zz,
    author = "Punturo, M. and others",
    editor = "Ricci, Fulvio",
    title = "{The Einstein Telescope: A third-generation gravitational wave observatory}",
    doi = "10.1088/0264-9381/27/19/194002",
    journal = "Class. Quant. Grav.",
    volume = "27",
    pages = "194002",
    year = "2010"
}

@article{Iacovelli:2022bbs,
    author = "Francesco Iacovelli and Michele Mancarella and Stefano Foffa and Michele Maggiore",
    title = "{Forecasting the Detection Capabilities of Third-generation Gravitational-wave Detectors Using GWFAST}",
    eprint = "2207.02771",
    archivePrefix = "arXiv",
    primaryClass = "gr-qc",
    doi = "10.3847/1538-4357/ac9cd4",
    journal = "Astrophys. J.",
    volume = "941",
    number = "2",
    pages = "208",
    year = "2022"
}

@article{Kalogera:2021bya,
    author = "Kalogera, Vicky and others",
    title = "{The Next Generation Global Gravitational Wave Observatory: The Science Book}",
    pages = "2111.06990",
    archivePrefix = "arXiv",
    primaryClass = "gr-qc",
    journal = {arXiv e-prints},
    month = "11",
    year = "2021"
}

@article{Baiotti:2005vi,
    author = "L. Baiotti and I. Hawke and L. Rezzolla and E. Schnetter",
    title = "{Gravitational-wave emission from rotating gravitational collapse in three dimensions}",
    eprint = "gr-qc/0503016",
    archivePrefix = "arXiv",
    doi = "10.1103/PhysRevLett.94.131101",
    journal = "Phys. Rev. Lett.",
    volume = "94",
    pages = "131101",
    year = "2005"
}

@article{LIGOScientific:2018mvr,
    title = "{GWTC-1: A Gravitational-Wave Transient Catalog of Compact Binary Mergers Observed by LIGO and Virgo during the First and Second Observing Runs}",
    author = {Abbott, B. P. and others},
    eprint = "1811.12907",
    archivePrefix = "arXiv",
    primaryClass = "astro-ph.HE",
    reportNumber = "LIGO-P1800307",
    doi = "10.1103/PhysRevX.9.031040",
    journal = "Phys. Rev. X",
    volume = "9",
    number = "3",
    pages = "031040",
    year = "2019"
}

@article{LIGOScientific:2017vwq,
    author = "Abbott, B. P. and others",
    collaboration = "LIGO Scientific, Virgo",
    title = "{GW170817: Observation of Gravitational Waves from a Binary Neutron Star Inspiral}",
    eprint = "1710.05832",
    archivePrefix = "arXiv",
    primaryClass = "gr-qc",
    reportNumber = "LIGO-P170817",
    doi = "10.1103/PhysRevLett.119.161101",
    journal = "Phys. Rev. Lett.",
    volume = "119",
    number = "16",
    pages = "161101",
    year = "2017"
}

@article{LIGOScientific:2021usb,
    author = "Abbott, R. and others",
    collaboration = "LIGO Scientific, VIRGO",
    title = "{GWTC-2.1: Deep extended catalog of compact binary coalescences observed by LIGO and Virgo during the first half of the third observing run}",
    eprint = "2108.01045",
    archivePrefix = "arXiv",
    primaryClass = "gr-qc",
    reportNumber = "LIGO-P2100063",
    doi = "10.1103/PhysRevD.109.022001",
    journal = "Phys. Rev. D",
    volume = "109",
    number = "2",
    pages = "022001",
    year = "2024"
}

@article{KAGRA:2021vkt,
    author = "Abbott, R. and others",
    collaboration = "KAGRA, VIRGO, LIGO Scientific",
    title = "{GWTC-3: Compact Binary Coalescences Observed by LIGO and Virgo during the Second Part of the Third Observing Run}",
    eprint = "2111.03606",
    archivePrefix = "arXiv",
    primaryClass = "gr-qc",
    reportNumber = "LIGO-P2000318",
    doi = "10.1103/PhysRevX.13.041039",
    journal = "Phys. Rev. X",
    volume = "13",
    number = "4",
    pages = "041039",
    year = "2023"
}

@article{Messick:2016aqy,
    author = "C. Messick and others",
    title = "{Analysis Framework for the Prompt Discovery of Compact Binary Mergers in Gravitational-wave Data}",
    eprint = "1604.04324",
    archivePrefix = "arXiv",
    primaryClass = "astro-ph.IM",
    doi = "10.1103/PhysRevD.95.042001",
    journal = "Phys. Rev. D",
    volume = "95",
    number = "4",
    pages = "042001",
    year = "2017"
}

@article{usman2016pycbc,
  title={The PyCBC search for gravitational waves from compact binary coalescence},
  author={S. A. Usman and others},
  journal={Classical and Quantum Gravity},
  volume={33},
  number={21},
  pages={215004},
  year={2016},
  publisher={IOP Publishing}
}

@article{Allen:2005fk,
    author = "Bruce Allen and Warren G. Anderson and Patrick R. Brady and Duncan A. Brown and Jolien D. E. Creighton",
    title = "{FINDCHIRP: An Algorithm for detection of gravitational waves from inspiraling compact binaries}",
    eprint = "gr-qc/0509116",
    archivePrefix = "arXiv",
    doi = "10.1103/PhysRevD.85.122006",
    journal = "Phys. Rev. D",
    volume = "85",
    pages = "122006",
    year = "2012"
}

@article{Sachdev:2019vvd,
    author = "Sachdev, S. and others",
    title = "{The GstLAL Search Analysis Methods for Compact Binary Mergers in Advanced LIGO's Second and Advanced Virgo's First Observing Runs}",
    pages = "1901.08580",
    archivePrefix = "arXiv",
    primaryClass = "gr-qc",
    month = "1",
    journal = {arXiv e-prints},
    year = "2019"
}

@article{Chaudhary:2023vec,
    author = "Chaudhary, S. S. and others",
    title = "{Low-latency gravitational wave alert products and their performance at the time of the fourth LIGO-Virgo-KAGRA observing run}",
    eprint = "2308.04545",
    archivePrefix = "arXiv",
    primaryClass = "astro-ph.HE",
    doi = "10.1073/pnas.2316474121",
    journal = "Proc. Nat. Acad. Sci.",
    volume = "121",
    number = "18",
    pages = "e2316474121",
    year = "2024"
}

@article{Kovalam_2022,
doi = {10.3847/2041-8213/ac5687},
url = {https://dx.doi.org/10.3847/2041-8213/ac5687},
year = {2022},
month = {mar},
publisher = {The American Astronomical Society},
volume = {927},
number = {1},
pages = {L9},
author = {Manoj Kovalam and Md Anwarul Kaium Patwary and Anala K. Sreekumar and Linqing Wen and Fiona H. Panther and Qi Chu},
title = {Early Warnings of Binary Neutron Star Coalescence Using the SPIIR Search},
journal = {The Astrophysical Journal Letters},
}

@article{Sachdev:2020lfd,
    author = "Sachdev, S. and others",
    title = "{An Early-warning System for Electromagnetic Follow-up of Gravitational-wave Events}",
    eprint = "2008.04288",
    archivePrefix = "arXiv",
    primaryClass = "astro-ph.HE",
    doi = "10.3847/2041-8213/abc753",
    journal = "Astrophys. J. Lett.",
    volume = "905",
    number = "2",
    pages = "L25",
    year = "2020"
}

@article{Nitz_2020,
doi = {10.3847/2041-8213/abbc10},
url = {https://dx.doi.org/10.3847/2041-8213/abbc10},
year = {2020},
month = {oct},
publisher = {The American Astronomical Society},
volume = {902},
number = {2},
pages = {L29},
author = {A. H. Nitz and M. Schäfer and T. Dal Canton},
title = {Gravitational-wave Merger Forecasting: Scenarios for the Early Detection and Localization of Compact-binary Mergers with Ground-based Observatories},
journal = {The Astrophysical Journal Letters},
}

@article{Wei:2020sfz,
    author = "Wei Wei and E. A. Huerta",
    title = "{Deep learning for gravitational wave forecasting of neutron star mergers}",
    eprint = "2010.09751",
    archivePrefix = "arXiv",
    primaryClass = "gr-qc",
    doi = "10.1016/j.physletb.2021.136185",
    journal = "Phys. Lett. B",
    volume = "816",
    pages = "136185",
    year = "2021"
}

@inproceedings{Baltus:2021emh,
  author={Grégory Baltus and Justin Janquart and Melissa Lopez and Amit Reza and Sarah Caudill and Jean-René Cudell},
  booktitle={2021 International Conference on Content-Based Multimedia Indexing (CBMI)}, 
  year={2021},
  volume={},
  number={},
  pages={1-6},
  doi={10.1109/CBMI50038.2021.9461919},
  title={}, }

@article{Wei:2020xrl,
    author = "Wei Wei and E. A. Huerta and Mengshen Yun and Nicholas Loutrel and Md Arif Shaikh and Prayush Kumar and Roland Haas and Volodymyr Kindratenko",
    title = "{Deep Learning with Quantized Neural Networks for Gravitational-wave Forecasting of Eccentric Compact Binary Coalescence}",
    eprint = "2012.03963",
    archivePrefix = "arXiv",
    primaryClass = "gr-qc",
    doi = "10.3847/1538-4357/ac1121",
    journal = "Astrophys. J.",
    volume = "919",
    number = "2",
    pages = "82",
    year = "2021"
}

@article{Alfaidi:2024ioo,
    author = "Reem Alfaidi and Christopher Messenger",
    title = "{Long Short-Term Memory for Early Warning Detection of Gravitational Waves}",
    pages = "2402.04589",
    archivePrefix = "arXiv",
    journal = {arXiv e-prints},
    primaryClass = "gr-qc",
    month = "2",
    year = "2024"
}

@article{Baltus:2022pep,
    author = "Grégory Baltus and Justin Janquart and Melissa Lopez and Harsh Narola and Jean-René Cudell",
    title = "{Convolutional neural network for gravitational-wave early alert: Going down in frequency}",
    eprint = "2205.04750",
    archivePrefix = "arXiv",
    primaryClass = "gr-qc",
    doi = "10.1103/PhysRevD.106.042002",
    journal = "Phys. Rev. D",
    volume = "106",
    number = "4",
    pages = "042002",
    year = "2022"
}

@article{2016arXiv160605328V,
       author = {Aaron van den Oord and Nal Kalchbrenner and Oriol Vinyals and Lasse Espeholt and Alex Graves and Koray Kavukcuoglu},
        title = "{Conditional Image Generation with PixelCNN Decoders}",
      journal = {arXiv e-prints},
         year = 2016,
        month = jun,
          eid = {arXiv:1606.05328},
        pages = {arXiv:1606.05328},
          doi = {10.48550/arXiv.1606.05328},
archivePrefix = {arXiv},
       eprint = {1606.05328},
 primaryClass = {cs.CV},
       adsurl = {https://ui.adsabs.harvard.edu/abs/2016arXiv160605328V}
}

@article{Baltus:2021nme,
    author = "Gregory Baltus and Justin Janquart and Melissa Lopez and Amit Reza and Sarah Caudill and Jean-Rene Cudell",
    title = "{Convolutional neural networks for the detection of the early inspiral of a gravitational-wave signal}",
    eprint = "2104.00594",
    archivePrefix = "arXiv",
    primaryClass = "gr-qc",
    reportNumber = "LIGO DCC number LIGO-P2100087",
    doi = "10.1103/PhysRevD.103.102003",
    journal = "Phys. Rev. D",
    volume = "103",
    pages = "102003",
    year = "2021"
}

@article{oord2016wavenet,
       author = {Aaron van den Oord and Sander Dieleman and Heiga Zen and Karen Simonyan and Oriol Vinyals and Alex Graves and Nal Kalchbrenner and Andrew Senior and Koray Kavukcuoglu},
        title = "{WaveNet: A Generative Model for Raw Audio}",
      journal = {arXiv e-prints},
         year = 2016,
        month = sep,
          eid = {arXiv:1609.03499},
        pages = {arXiv:1609.03499},
          doi = {10.48550/arXiv.1609.03499},
archivePrefix = {arXiv},
       eprint = {1609.03499},
 primaryClass = {cs.SD},
       adsurl = {https://ui.adsabs.harvard.edu/abs/2016arXiv160903499V}
}

@inproceedings{Que:2021cqo,
    author = "Que, Zhiqiang and others",
    booktitle = "{32nd IEEE International Conference on Application-specific Systems, Architectures and Processors}",
    eprint = "2106.14089",
    archivePrefix = "arXiv",
    primaryClass = "cs.LG",
    doi = "10.1109/ASAP52443.2021.00025",
    month = "6",
    year = "2021",
    title = {},
}

@article{2020arXiv200204591C,
       author = {Marco Cavaglia and Sergio Gaudio and Travis Hansen and Kai Staats and Marek Szczepanczyk and Michele Zanolin},
        title = "{Improving the background of gravitational-wave searches for core collapse supernovae: A machine learning approach}",
      journal = {arXiv e-prints},
         year = 2020,
        month = feb,
          eid = {arXiv:2002.04591},
        pages = {2002.04591},
          doi = {10.48550/arXiv.2002.04591},
archivePrefix = {arXiv},
       eprint = {2002.04591},
 primaryClass = {astro-ph.IM},
       adsurl = {https://ui.adsabs.harvard.edu/abs/2020arXiv200204591C}
}

@article{2015arXiv151203385H,
       author = {Kaiming He and Xiangyu Zhang and Shaoqing Ren and Jian Sun},
        title = "{Deep Residual Learning for Image Recognition}",
      journal = {arXiv e-prints},
         year = 2015,
        month = dec,
          eid = {arXiv:1512.03385},
        pages = {1512.03385},
          doi = {10.48550/arXiv.1512.03385},
archivePrefix = {arXiv},
       eprint = {1512.03385},
 primaryClass = {cs.CV},
       adsurl = {https://ui.adsabs.harvard.edu/abs/2015arXiv151203385H}
}

@article{Boudart:2022xib,
    author = "V. Boudart and M. Fays",
    title = "{Machine learning algorithm for minute-long burst searches}",
    eprint = "2201.08727",
    archivePrefix = "arXiv",
    primaryClass = "gr-qc",
    reportNumber = "4126596",
    doi = "10.1103/PhysRevD.105.083007",
    journal = "Phys. Rev. D",
    volume = "105",
    number = "8",
    pages = "083007",
    year = "2022"
}

@article{Boudart:2022apz,
    author = "V. Boudart",
    title = "{Convolutional neural network to distinguish glitches from minute-long gravitational wave transients}",
    eprint = "2210.04588",
    archivePrefix = "arXiv",
    primaryClass = "gr-qc",
    doi = "10.1103/PhysRevD.107.024007",
    journal = "Phys. Rev. D",
    volume = "107",
    number = "2",
    pages = "024007",
    year = "2023"
}

@article{bahaadini2018machine,
title = {Machine learning for Gravity Spy: Glitch classification and dataset},
journal = {Information Sciences},
volume = {444},
pages = {172-186},
year = {2018},
issn = {0020-0255},
doi = {https://doi.org/10.1016/j.ins.2018.02.068},
url = {https://www.sciencedirect.com/science/article/pii/S0020025518301634},
author = {S. Bahaadini and V. Noroozi and N. Rohani and S. Coughlin and M. Zevin and J.R. Smith and V. Kalogera and A. Katsaggelos},
}

@article{Laguarta:2023evo,
   author = {P. {Laguarta} and others},
        title = "{Detection of anomalies amongst LIGO's glitch populations with autoencoders}",
      journal = {Classical and Quantum Gravity},
         year = 2024,
        month = {mar},
       volume = {41},
       number = {5},
          eid = {055004},
        pages = {055004},
          doi = {10.1088/1361-6382/ad1f26},
archivePrefix = {arXiv},
       eprint = {2310.03453},
}

@inproceedings{Dooney:2022arh,
author = {Tom Dooney and Stefano Bromuri and Lyana Curier},
booktitle = {2022 IEEE International Conference on Big Data (Big Data)},
year = {2022},
volume = {},
issn = {},
pages = {5468-5477},
doi = {10.1109/BigData55660.2022.10021080},
publisher = {IEEE Computer Society},
address = {Los Alamitos, CA, USA},
month = {dec},
title = {},
}

@article{Lopez:2022lkd,
    author = "Melissa Lopez and Vincent Boudart and Kerwin Buijsman and Amit Reza and Sarah Caudill",
    title = "{Simulating transient noise bursts in LIGO with generative adversarial networks}",
    eprint = "2203.06494",
    archivePrefix = "arXiv",
    primaryClass = "astro-ph.IM",
    doi = "10.1103/PhysRevD.106.023027",
    journal = "Phys. Rev. D",
    volume = "106",
    number = "2",
    pages = "023027",
    year = "2022"
}

@article{Lopez:2022dho,
    author = "Melissa Lopez and Vincent Boudart and Stefano Schmidt and Sarah Caudill",
    title = "{Simulating Transient Noise Bursts in LIGO with gengli}",
    pages = "2205.09204",
    archivePrefix = "arXiv",
    primaryClass = "astro-ph.IM",
    month = "5",
    year = "2022",
    journal = "arXiv pre-print"
}

@article{LopezPortilla:2020odz,
    author = "M. Lopez Portilla and I. Di Palma and M. Drago and P. Cerda-Duran and F. Ricci",
    title = "{Deep learning for core-collapse supernova detection}",
    eprint = "2011.13733",
    archivePrefix = "arXiv",
    primaryClass = "astro-ph.IM",
    doi = "10.1103/PhysRevD.103.063011",
    journal = "Phys. Rev. D",
    volume = "103",
    number = "6",
    pages = "063011",
    year = "2021"
}

@inproceedings{Lopez:2021rci,
  author={M. López and M. Drago and I. Di Palma and F. Ricci and P. Cerdá-Durán},
  booktitle={2021 International Conference on Content-Based Multimedia Indexing (CBMI)}, 
  year={2021},
  volume={},
  number={},
  pages={1-6},
  doi={10.1109/CBMI50038.2021.9461885},
title={}, 
}

@article{Cuoco:2020ogp,
    author = "E. Cuoco and others",
    title = "{Enhancing Gravitational-Wave Science with Machine Learning}",
    eprint = "2005.03745",
    archivePrefix = "arXiv",
    primaryClass = "astro-ph.HE",
    doi = "10.1088/2632-2153/abb93a",
    journal = "Mach. Learn. Sci. Tech.",
    volume = "2",
    number = "1",
    pages = "011002",
    year = "2021"
}

@article{vanStraalen:2024xiq,
    author = "Wouter van Straalen and Alex Kolmus and Justin Janquart and Chris Van Den Broeck",
    title = "{Pre-Merger Detection and Characterization of Inspiraling Binary Neutron Stars Derived from Neural Posterior Estimation}",
    journal = {arXiv e-prints},
    pages = "2407.10263",
    archivePrefix = "arXiv",
    primaryClass = "gr-qc",
    month = "7",
    year = "2024"
}

@INPROCEEDINGS{6589302,
  author={Liu, Bin and Zydek, Dawid and Selvaraj, Henry and Gewali, Laxmi},
  booktitle={2012 13th International Conference on Parallel and Distributed Computing, Applications and Technologies}, 
  year={2012},
  volume={},
  number={},
  pages={337-342},
  doi={10.1109/PDCAT.2012.34},
 title={}, 
}

@article{amaroseoane2017laserinterferometerspaceantenna,
      title={Laser Interferometer Space Antenna}, 
      author={P. Amaro-Seoane and others},
      year={2017},
      pages={1702.00786},
      archivePrefix={arXiv},
      primaryClass={astro-ph.IM},
      url={https://arxiv.org/abs/1702.00786},
      journal = {arXiv e-prints}
}

@article{andre2018bigdata,
title={{Big data and extreme-scale computing : Pathways to Convergence-Toward a shaping strategy for a future software and data ecosystem for scientific inquiry}},
  author={Asch, M. and others},
  url ={https://hal.science/hal-03800621},
  journal={{International Journal of High Performance Computing Applications}},
  publisher={{SAGE Publications}},
  volume={32},
  number={4},
  pages={435-479},
  year={2018},
  month={Jul},
  doi={10.1177/1094342018778123},
}

@INPROCEEDINGS{9407142,
  author={Udit Gupta and Young Geun Kim and Sylvia Lee and Jordan Tse and Hsien-Hsin S. Lee and Gu-Yeon Wei and David Brooks and Carole-Jean Wu},
  booktitle={2021 IEEE International Symposium on High-Performance Computer Architecture (HPCA)}, 
  title={Chasing Carbon: The Elusive Environmental Footprint of Computing}, 
  year={2021},
  volume={},
  number={},
  pages={854-867},
  doi={10.1109/HPCA51647.2021.00076},
title={}, 
}

@article{Aveiro_2022,
   title={Identification of binary neutron star mergers in gravitational-wave data using object-detection machine learning models},
   volume={106},
   ISSN={2470-0029},
   url={http://dx.doi.org/10.1103/PhysRevD.106.084059},
   DOI={10.1103/physrevd.106.084059},
   number={8},
   journal={Physical Review D},
   publisher={American Physical Society (APS)},
   author={João Aveiro and Felipe F. Freitas and Márcio Ferreira and Antonio Onofre and Constança Providência and Gonçalo Gonçalves and José A. Font},
   year={2022},
   month=oct }

@article{miyashita2016convolutionalneuralnetworksusing,
      title={Convolutional Neural Networks using Logarithmic Data Representation}, 
      author={Daisuke Miyashita and Edward H. Lee and Boris Murmann},
      year={2016},
      pages={1603.01025},
      archivePrefix={arXiv},
      primaryClass={cs.NE},
      url={https://arxiv.org/abs/1603.01025}, 
    journal = {arXiv e-prints},
}

\end{document}